\documentstyle[prb,aps,eqsecnum,epsf,twocolumn]{revtex}
\newcommand{\beq}{\begin{equation}}
\newcommand{\eeq}{\end{equation}}
\newcommand{\bea}{\begin{eqnarray}}
\newcommand{\eea}{\end{eqnarray}}

\newcommand{\etal}{{\em et al.}}

\def\bv#1{{\bf v}_{#1}}
\def\expp#1{e^{#1}}
\def\expP#1#2{\exp(\fracc{#1\pi i}{#2})}

\def\jour#1#2#3#4{{#1}{\bf #2}, #3 (#4)}
\def\tit#1#2#3#4#5{{#1}{\bf #2}, #3 (#4)}
\def\jmp{J.\ Math.\ Phys.\ }

\def\npb{Nucl.\ Phys.\ B\ }

\def\prl{Phys.\ Rev.\ Lett.\ }
\def\pr{Phys.\ Rev.\ }
\def\prb{Phys.\ Rev.\ B\ }

\def\zpb{Z.\ Phys.\ B\ }
\def\jpsj{J.\ Phys.\ Soc.\ Jpn.\ }
\def\sci{Science\ }

\def\epjb{Eur.\ Phys.\ J.\ B\ }
\def\pla{Phys.\ Lett.\ A\ }

\def\fracc#1#2{{#1}/{#2}}

\begin{document}
\draft

\twocolumn[\hsize\textwidth\columnwidth\hsize\csname @twocolumnfalse\endcsname

\title{ Ising models of quantum frustration}
 
\author{R. Moessner and S. L. Sondhi}

\address{Department of Physics, Princeton University,
Princeton, NJ 08544, USA }
\date{\today}

\maketitle

\begin{abstract}
We report on a systematic study of two dimensional, periodic,
frustrated Ising models with a quantum dynamics introduced via a
transverse magnetic field. The systems studied are the triangular and
kagome lattice antiferromagnets, fully frustrated models on the square
and hexagonal (honeycomb) lattices, a planar analog of the pyrochlore
antiferromagnet, a pentagonal lattice antiferromagnet as well as a two
quasi one-dimensional lattices that have considerable pedagogical
value. All of these exhibit a macroscopic degeneracy at $T=0$ in the
absence of the transverse field, which enters as a singular
perturbation. We analyze these systems with a combination of a
variational method at weak fields, a perturbative
Landau-Ginzburg-Wilson (LGW) approach from large fields as well as
quantum Monte Carlo simulations utilizing a cluster algorithm. Our
results include instances of quantum order arising from classical
criticality (triangular lattice) or classical disorder (pentagonal and
probably hexagonal) as well as notable instances of quantum disorder
arising from classical disorder (kagome).  We also discuss the effect
of a finite temperature, as well as the interplay between longitudinal
and transverse fields---in the kagome problem the latter gives rise to
a non-trivial phase diagram with bond-ordered and bond-critical phases
in addition to the disordered phase. We also note connections to
quantum dimer models and thereby to the physics of Heisenberg
antiferromagnets in short-ranged resonating valence bond phases that
have been invoked in discussions of high-temperature
superconductivity.
\end{abstract}


]

\section{Introduction}

\begin{figure}
\epsfxsize=1.2in
\centerline{\epsffile{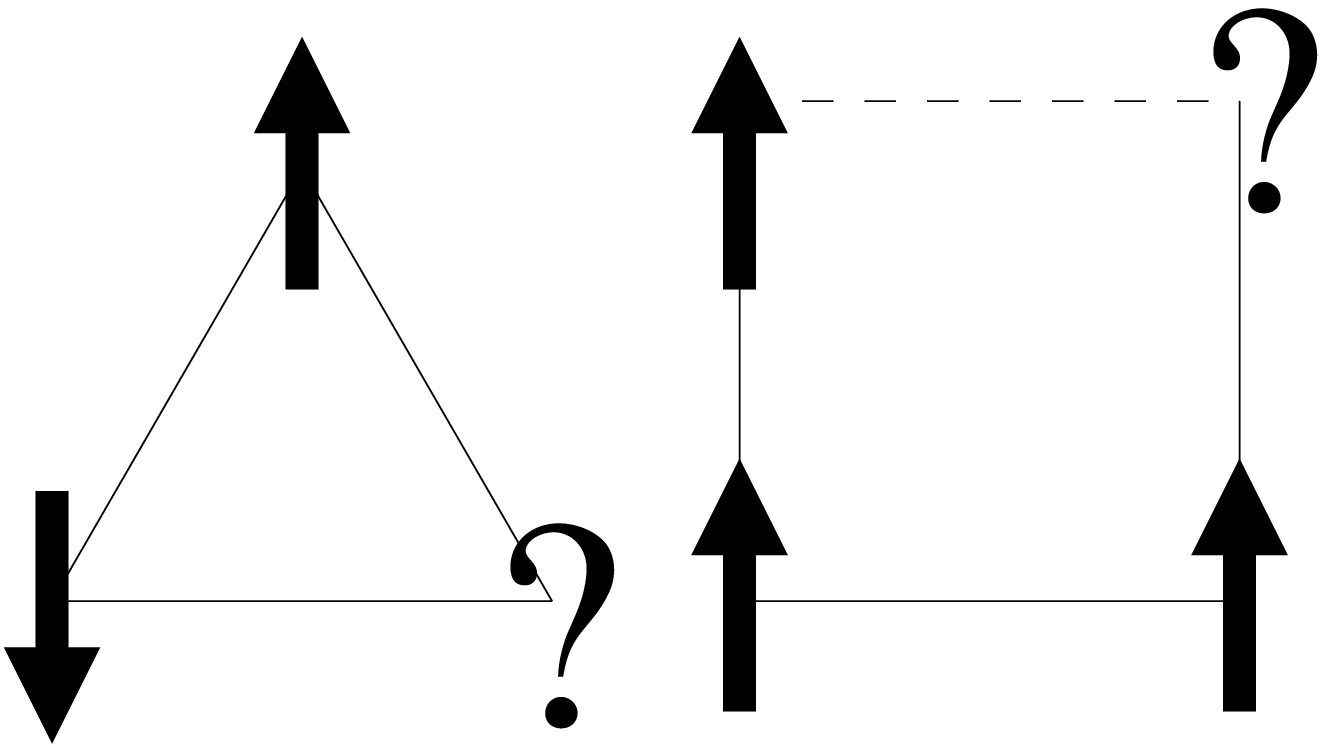}}
\caption{ Two frustrated plaquettes with Ising spins. In the canonical
example of the antiferromagnetic triangle, anti-aligning two spins
leaves the direction of the third undetermined. Similarly, a square
can be frustrated by choosing an odd number of bonds to be
antiferromagnetic. For such mixed-bond models, we represent
(anti)ferromagnetic bonds by (dashed) solid lines.}
\label{fig:frusplaq}
\end{figure}

The study of frustrated magnetic systems began a half century ago with
the realization by Wannier\cite{wannier} and Houtappel\cite{houtappel}
that the antiferromagnetic Ising model on a two-dimensional triangular
lattice does not order down to the lowest temperatures and exhibits a
finite entropy density even at $T=0$, in contrast to the naive
expectation from the third law of thermodynamics. These two
observations are related, and can be traced to the frustrated nature
of the couplings even at the level of a single plaquette
(Fig.~\ref{fig:frusplaq}) where one sees that it is not possible to
minimize the energy of all three bonds simultaneously leading to six
(instead of the two for ferromagnets) minimum energy
configurations. From this we may infer a macroscopic entropy density
for the triangular lattice at $T=0$ and rationalize the absence of any
ordering by the easy accessibility of a large number of configuration
at all temperatures.\cite{fn-caveat1}

The twin observations, of a non-vanishing entropy and lack of order,
when contrasted with the ordering transition and small number (two) of
ground states in the ferromagnetic Ising model on the same lattice,
typify the striking behavior of such ``maximally'' frustrated
classical models whose catalog, by now, includes also models with
continous spins such as the Heisenberg magnet on the three dimensional
pyrochlore lattice.\cite{villain,hfmrev} Other, by definition, less
frustrated models exhibit non-obvious phase transitions\cite{fn-frus}
at finite temperatures. These may involve a singular privileging of
the ground state manifold as a whole as in the case of the XY magnet
on the kagome lattice\cite{Huse92} or, more typically, the elegant
phenomenon of ``order by disorder'' \cite{Villain80,shenderquantum} in
which ground states that permit especially soft fluctuations about
themselves are selected entropically at finite temperatures. 

This paper is concerned with the fate of such frustrated models,
especially the maximally frustrated ones, when they are endowed with a
quantum dynamics at $T=0$. The canonical problems of this kind are
$S=1/2$ antiferromagnetic quantum Heisenberg models on
triangular\cite{collinsrev} and kagome lattices\cite{kagdiag} which
have both attracted a considerable amount of interest. While one can
think of these as quantized versions of classical Heisenberg models,
it is instructive instead to follow Anderson and
Fazekas\cite{Fazekas74} and think of them as perturbed variants of
their Ising limits, which are manifestly frustrated. In these
particular cases, the perturbation is the XY exchange and as it
does not commute with the Ising pieces, it introduces a quantum
(two-spin) dynamics into the frustrated problem. This perspective, in
turn, suggests consideration of a more general class of models with
alternative perturbations introducing a quantum dynamics instead.

In this paper we report on a systematic study of frustrated Ising
models perturbed by transverse fields -- as these introduce a single
spin dynamics, they are evidently the simplest models in the extended
class.\cite{fn-extensions} A short, partial, account of this work has
appeared previously.\cite{mcs2000} This paper covers a broader range
of issues (and lattices) and supplies many of the details left out in
the short account. It also provides a compendium of results we have
obtained in the course of our work, which we hope will be of some use
to people undertaking further study in this field.  It is perhaps
worth noting that this simplicity does make the models more tractable,
which is generally true for transverse field Ising models and accounts
for their ubiquity in quantum statistical mechanical contexts. Much of
this work is reviewed in Ref.~\onlinecite{Chakrabarti96}, which also
includes some previous work on one-dimensional frustrated chains. More
recent studies include the use of the unfrustrated model as a paradigm
of quantum critical behavior\cite{Sachdev99} and the treatment of
random versions by asymptotically exact real space renormalization
techniques near infinite disorder fixed points.\cite{rsrg}

A second motivation for studying these models is the possibility of
direct experimental realization. Ising systems exist and in the case
of LiHoF$_4$ and its kin, transverse fields have been used to tune
between phases in clean and disordered systems.\cite{Aeppli98} A
second family of Ising systems consists of stacked triangular
lattices, strongly coupled along the stack, reviewed at length in
Ref.~\onlinecite{collinsrev}. While these have an intimate connection
to the single triangular lattice in a transverse field via its
euclidean representation, it would be interesting in future work to
consider the effect of a transverse field on them. Meanwhile we would
encourage our experimental colleagues to search for a triangular or
kagome Ising antiferromagnet which are both, as we shall show in this
paper, exemplars of very different physics that can arise in
frustrated quantum systems.

Finally, there is the possibility of finding systems where there is a
local Ising degree of freedom that is not itself an Ising spin. One
such connection, which is currently the subject of intense interest,
is an exact mapping from frustrated transverse field Ising
models\cite{mcs2000} to quantum dimer models of the short-range
resonating valence bond state conjectured a while back by
Anderson\cite{pwa87} and oft mentioned in discussions of the cuprate
superconductors; in this context the Ising models appear more
naturally as their dual, Ising gauge theories.
\cite{Rokhsar88,readsach,wen,mudry94,SV,SF,MStrirvb}

Turning now to the physics of the models themselves, we note that
the introduction of quantum fluctuations can be expected to lead to a
variety of behaviors much as in the case of thermal fluctutations
catalogued above. Indeed, in the quantum case, the singular character
of the fluctuations is manifest in that even infinitesimal strength
perturbations lead to a non-trivial problem of degenerate perturbation
theory in a macroscopically degenerate manifold and can therefore be
expected to lift the degeneracy and select a much smaller number of
ground states. Consequently there must be a discontinuity in the
entropy and ground state correlations at zero quantum coupling and
$T=0$. We should emphasize that degenerate perturbation theory
problems are not themselves perturbative, especially for macroscopic
degeneracies. The most notable example of this is the quantum Hall
problem where the degeneracy of a partially filled Landau level is
lifted by the interaction and disorder in ways that lead to an
incredibly complex phase diagram. Part of the interest of studying a
diversity of perturbations of frustrated magnets is the prospect of
generating at least a fraction of this complexity.

Two possibilities are generic at small quantum couplings: a) a quantum
version of order by disorder\cite{Villain80,shenderquantum} in which a
broken symmetry state constructed out the the degenerate manifold is
selected and b) a quantum version of the disordered or (cooperative)
paramagnetic possibility in which the ground state correlations remain
short ranged and might be expected to lead to the opening of a gap in
the spectrum. This latter possibility, which we have christened
``disorder by disorder'', was first suggested by Anderson and
Fazekas\cite{Fazekas74} as a mechanism for obtaining spin-liquid
states of the RVB kind. Further singularities can emerge at large
quantum couplings. For single spin dynamics, such as the transverse
field problems we study, there is neccessarily a paramagnetic phase at
large couplings and at least one phase transition {\sl en route} in
cases of order by disorder. The nature of such phase transitions,
which will be of some interest to us in this paper, can indeed be very
unconventional.

Specifically, we study Hamiltonians of the form: 
\beq H = \sum_{\langle ij \rangle} J_{ij} S^z_i
S^z_j + \Gamma \sum_{i} S^x_i + h \sum_{i} S^z_i \ ,
\label{eq:hamil}
\eeq on a variety of one and two dimensional lattices (see
Fig.~\ref{fig:alllattices}). Here the $J_{ij}$ are nearest-neighbor
exchange couplings with $|J_{ij}| = J$ and ${\prod_{\rm plaquette}
(\fracc{-J_{ij}}{J}}) = -1$, $\Gamma$ is the strength of the
transverse field, the $S^a$ are the Pauli spin operators and $h$ is
the strength of a (classical) longitudinal, Ising symmetry breaking
field.

The structure of this paper is as follows. We first
introduce concepts and methods we have found useful in studying the
model under consideration (Sects.~\ref{sect:methods} and
\ref{sect:tll}). These we apply in the remainder of the paper to a
number of frustrated Ising models on different lattices which, between
them, realise a wide range of classical and quantum properties. We
conclude by discussing possible realisations of such models. 

In detail, in Sect.~\ref{sect:methods}, we use a mapping of the $d$
dimensional quantum model to a $d+1$ dimensional classical model to
derive a general criterion for the existence of a quantum ordering
transition, and we use this mapping to obtain a quantum Monte Carlo
algorithm.  This algorithm is free of any sign problems and is one of
the attractions of studying this class of models -- augmented by a
cluster method appropriate for our problems, we have used it at
several points in this work.

In Sect.~\ref{sect:tll}, we consider a three-leg ladder that has
considerable pedagogical value and allows us to introduce a weak
coupling ($\Gamma/J \ll 1$) variational analysis and a strong coupling
Landau-Ginzburg-Wilson analysis that will be our principal analytic
tools in the remainder of the paper. Sect.~\ref{sect:pent} reports
results on the somewhat baroque pentagonal lattice antiferromagnet
which turns out to have an intimate connection with the ladder
described previously. Its classically disordered state gives way to
quantum order.  In the next three sections (Sect.~\ref{sect:triang},
\ref{sect:villain}, \ref{sect:twodpyro}) we discuss three models: the
triangular lattice antiferromagnet, the fully frustrated square
lattice and the ``two dimensional pyrochlore'' lattice (see
Fig.~\ref{fig:alllattices}), respectively.  These exhibit critical
classical correlations in their ground state manifolds, and undergo
ordering transitions, in accordance with the ordering criterion
derived in Sect.~\ref{sect:methods}. The latter also exhibits two
unconventional critical phases.

Next we turn to classically strongly disordered
systems. Sect.~\ref{sect:kagome} deals with the kagome lattice
antiferromagnet which is a notable instance of disorder by disorder
and also exhibits a highly non-trivial phase diagram when both
transverse and longitudinal fields are
present. Sect.~\ref{sect:sawtooth} deals with a one dimensional Ising
quantum disordered magnet, the sawtooth chain, the classical version
of which is the ultimate cooperative paramagnet.  Finally,
Sect.~\ref{sect:ffhexagon} analyzes the fully frustrated honeycomb
lattice which appears to exhibit a fairly complex pattern of ordering
driven by quantum fluctuations as well as a non-trivial, $O(4)$ phase
transition. 

At various places in the paper we discuss connections to quantum
height or dimer models.  Via the latter, we find a connection to
frustrated, valence bond, phases of {\it Heisenberg} magnets that has
been of interest starting from the oppposite
end\cite{readsach,wen,mudry94} and more recently starting from d-wave
superconductors.\cite{SF,SV} As noted before, we are hopeful that this 
represents a more general possibility of realizing the models we study here in
other contexts where frustration is present and it is possible to
focus on a local Ising degree of freedom. We close with a brief 
recapitulation of our themes in the summary section.

\section{A criterion for ordering and a method for quantum Monte Carlo}
\label{sect:methods}

In this section we map our $d$-dimensional quantum magnet onto a
$d+1$-dimensional ferromagnetically stacked, classical magnet. This
mapping allows use to derive some qualitative features of the quantum
ordering process, and also to make contact with the existing
literature on stacked magnets. In addition, it enables us to devise a
quantum Monte Carlo code free of the sign problem for general
classical terms in the Hamiltonian.

\subsection{Mapping onto the stacked magnet}
\label{subsect:mapstack}
We~proceed~by~using the Suzuki-Trotter
formalism,\cite{Suzuki71,Trotter59} to determine the precise
correspondence between the quantum $d$\ and classical $d+1$
dimensional models.  Specifically, the partition function of the
transverse-field Ising model 
is
\bea 
Z &=& {\rm Tr}e^{-\beta H} = {\rm Tr} \exp \left\{-\beta 
\left(
\upsilon V(\{ S_i^z\}) + \Gamma \sum_{i} S^x_i
\right)
\right\}
\label{eq:part}
\eea where $\beta \equiv \fracc{1}{k_B T}$.  For generality, we have
introduced the notation $V(\{S_i^z\})$\ for the `classical' 
part of the
Hamiltonian of strength $\upsilon$,
 i.e., $[V,S_i^z]=0$ for all $i$; in the simplest case,
$V$\ only consists of the exchange part of $H$, $\upsilon
V=\sum_{\langle ij
\rangle} J_{ij} S_i^z S^z_j$.

We now use a path integral
representation
of (\ref{eq:part}) where the insertion of a complete
set of states effectively introduces an additional dimension
of size $\beta$,
we follow Suzuki's
approach\cite{Suzuki71}:
\bea
%Z&=&\sum_{\{S_i\}}\langle\{S_i\}\left|\exp(-\beta H)\right|\{S_i\}
%\rangle\nonumber\\
Z&=&\sum_{\{S_i\}}\langle\{S_i\}\left|(\exp(-a_\tau H))^N\right|\{S_i\}\rangle
\nonumber \\
&=&\sum_{n=1}^{\beta/a_\tau}\sum_{\{S_{i,n}\}}
\langle\{S_{i,n}\}\left|\exp(-a_\tau H)\right|\{S_{i,n+1}\}\rangle\\
&=&\sum_{n=1}^{\beta/a_\tau}\sum_{\{S_{i,n}\}}
\exp\left(-a_\tau\upsilon V(\{S_{i,n}\}) \right)\times\\
&&\times
\left( 
\delta^{(0)}_{\{S_{i,n}\},\{S_{i,n+1}\}}+
\frac{1}{2}a_\tau\Gamma\delta^{(1)}_{\{S_{i,n}\},\{S_{i,n+1}\}}+
O((a_\tau\Gamma)^2)\nonumber
\right)
%\langle\{S_{i,n}\}\left|\exp(-\beta H)\right|\{S_{i,n}\}\rangle ;
\label{eq:expand}
\eea 
Here, we have introduced the imaginary time step $a_\tau$\ and
$n$\ labels the coordinate of the extra dimension. The function 
$\delta^{(k)}$\ is defined to be one if its arguments, 
the two (ordered) sets of spin configurations differ by $k$\ entries, 
and zero otherwise. 

We now establish an equivalence between
the transverse field model and the classical Hamiltonian  
\bea
H_{d+1} &=& \sum_{\langle ij \rangle, n} 
%K^s_{ij} S_{i,n}S_{j,n}
K^s_V V(\{S_{i,n}\})
  +\sum_{i,n} K^\tau S_{i,n} S_{i,n+1}
\label{eq:cham}
\eea 
by expressing the partition sum for the latter in terms of a 
transfer matrix, $T_Z$:
\bea
T_Z&=&\exp\left(K^s_V V(\{S_{i,n}\})   \right)\times
\nonumber\\
&&\times\left(\delta^{(0)}_{\{S_{i,n}\},\{S_{i,n+1}\}}+
\exp\left(-K^\tau/2 \right)
\delta^{(1)}_{\{S_{i,n}\},\{S_{i,n+1}\}}+\right.
\nonumber\\
%O((a_\tau\Gamma)^2)
%\exp\left(-K^\tau/2 \right)
&&+ \left.O\left(\exp\left(-K^\tau \right)\right)
\right)\ .
\eea
The first term on the right hand side of the previous equation
is to be understood as an exponentiated diagonal matrix.

>From this, one can read off that the two partition functions will be
equivalent if one chooses $a_\tau\Gamma/2=\exp(-K^\tau/2)$\ and
$K^s_V=a_\tau\upsilon$.  We note that continuous quantum evolution
corresponds to the scaling 
limit $K^s_V\propto a_\tau \rightarrow 0$, $K^\tau
\rightarrow \infty$, while maintaining
\bea
2e^{-K^\tau/2}/a_\tau=\Gamma, \ \ \ 
K^s_V\exp(K^\tau/2)/2=\upsilon/\Gamma\ .
\eea

With the classical Hamiltonian $H_{d+1}$ (Eq.~\ref{eq:cham}), the
Ising spins interact in the spatial layers as they do in the analogous
$d$ dimensional classical problem, but they are also coupled {\sl
ferromagnetically} in the additional (imaginary time) dimension. The
dimensionless inverse of the quantum temperature, $\beta \Gamma$, is
given by the extent $L^\tau$\ of the system in the time direction:
$\beta\Gamma=a_\tau\Gamma L^\tau= 2\exp(-K^\tau/2)L^\tau$.

\subsection{A criterion for order by disorder}
\label{subsect:criterion}

We can look at the possibility of quantum ordering in a transverse
field by studying the discretized partition function $Z_\infty(K^\tau)
\equiv Z(K^s \rightarrow \infty, K^\tau)$ as an expansion in powers of
$K^\tau$. In this limit,
we force the spin configurations in each plane
%$\{S_i^{gs}\}$
to be classical ground states. Taking the trace, ${\rm Tr_{gs}}$\ over
 these ground states gives
\beq
Z_{\infty}(K^\tau) = {\rm Tr_{gs}} \exp\left( K^\tau 
\sum_{i,n} S_{i,n} S_{i,n+1} \right)
\label{eq:zkinfinite}
\eeq
where the sum is over all ground-states for each layer.  Expanding in
small $K^\tau$, we obtain 
\bea
Z_{\infty}(K^\tau) &=& {\rm Tr_{gs}} \left(1 + \frac{1}{2} (K^\tau)^2
\times\label{eq:zseries}
\right.\\
&&\left.
\times
\sum_{i,n} \sum_{j,n'} S_{i,n} S_{i,n+1}
S_{j,n^\prime}S_{j,n^\prime+1} + O((K^\tau)^4) \right)
\nonumber
%\label{eq:zseries}
\eea
where the linear term in $K^\tau$ is absent due to Ising symmetry.  We
can further express Eq.~(\ref{eq:zseries}) as
\bea
Z_{\infty}(K^\tau) &=& Z(L)^{L^\tau} \left\{1 + 
\frac{1}{2} (K^\tau)^2 \sum_{i,n} \langle S_{i,n}
S_{i,n+1}\rangle^2 + ... \right\}
\label{eq:zseries2}
\eea
where the prefactor refers to the number of classical states in each
layer of area $L^2$ and $L^\tau \equiv \beta/a_\tau$ is the number of
layers.  
Then
\bea
Z_\infty(K^\tau) &=& \exp \left(L^2 L^\tau {\cal S}\right)
\times
\label{eq:zseries3}
\\ 
&&\times
\left\{1 + \frac{1}{2}(K^\tau)^2 (L^2L^\tau) 
\sum_{i} \langle 
S_{i,0} S_{0,0}
\rangle ^2 + ...  \right\}
%\label{eq:zseries3}
\nonumber 
\eea 
where ${\cal S}$ is the classical ground-state entropy
density.
Therefore the effective free energy, 
$F_\infty=- \ln Z_\infty/\beta$\, as a function
of small $K^\tau$ is
\beq
-\beta F_\infty = (L^2 L^\tau) \left\{{\cal S} + 
\frac{(K^\tau)^2}{2}
\sum_{i} \langle S_{i,0} S_{0,0} \rangle^2 + ... \right\}
\label{eq:free}
\eeq
where the term $(L^2 L^\tau)$ is a volume in space-time.  

This yields the following powerful result.  If the sum 
$I=\sum_{i}\langle S_{i,0} S_{0,0} \rangle^2$\ 
%$I = \int d^2 r \langle S(0) S(r) \rangle^2$\ 
of the 
{\em classical} correlation function  
diverges, the free energy above is {\sl non-analytic} as 
$K^\tau \rightarrow
0$, implying that the $K_{ij}^s = \infty$ is in
a different phase than the disordered point $K^\tau =0$ for {\sl any}
$K^\tau > 0$; this is indeed the case for the Ising triangular
antiferromagnet where $I_{tri}$\ contains a leading divergence 
of the form\cite{stephensontri}
$\int d^2 r
\left(\fracc{1}{\sqrt{r}}\right)^2$. In this case, one does
indeed find
quantum order by disorder (see Sect.~\ref{sect:triang}).

However the situation is inconclusive for classically disordered
antiferromagnets with exponentially decaying spin correlations at
$T=0$; even though no single term in the series expansion,
Eq.~(\ref{eq:free}), becomes unbounded, the whole series may diverge
because it is beyond its radius of convergence.  
There is therefore no clear link between classical and quantum
disorder.

\subsection{Quantum Monte Carlo}
By simulating the stacked {\em classical} magnet, it is thus possible
to gain information on the properties of the quantum system. We note
that, thanks to the simplicity of our quantum dynamics, there is no
sign problem to cope with for this class of models. 

However, the scaling limit $K^\tau\rightarrow\infty$\ does pose some
technical problems. At large $K^\tau$, domain walls in the time
direction become very rare, which leads to a divergent timescale in
the Monte Carlo simulations. This problem can be remedied by employing
a cluster algorithm, in which the attempted Monte Carlo moves consist
not of flipping a single spin but rather rods of spins in the time
direction. These rods can be chosen in a way which exactly cancels the
inclement Boltzmann factor $\exp(-K^\tau/2)$.\cite{ferswen}  
It is with this method
that we have carried out the simulations presented here, with an
additional feature for the fully frustrated hexagonal lattice, which
is described there (see Sect.~\ref{sect:ffhexagon}).

It however turns out that the dominant source of error can be the
remaining discretisation error. To see this, consider the case of a
quantum ordered state at zero quantum temperature, $1/\beta=0$, which
corresponds to an infinite extension of the stacked model in the time
direction.  One can now imagine starting at an effectively infinite
classical coupling, $K^\tau$. As $K^\tau$\ is reduced, within the
framework of the above mapping, one retains $1/\beta=0$\ but a
discretisation error is introduced -- which for sufficiently small
$K^\tau$\ will be serious enough to make the quantum order disappear
as the classical magnet goes through its transition at a critical
$K^\tau$. If one considers a finite $\beta$, one in practice has to
trade off this error, which requires large $K^\tau$\, against reaching
a low quantum temperature, which nominally decreases with increasing
$K^\tau$.

To quantify the discretisation error, we quote the parameter
$\lambda\equiv \exp(k K^\tau/2)$, (where $k$\ is the multiplicity of
the spin flip, being one for single-spin dynamics), a lengthscale
characteristic of an isolated ferromagnetically coupled rod -- the
larger $\lambda$, keeping $\lambda/L^\tau$\ fixed, the better. For
fixed $L^\tau$, however, the optimal value of $\lambda$ depends on the
correlation length in the time direction, which varies from system to
system and which is larger the smaller the quantum gap.

In addition, the time for building up the clusters also becomes large
at low temperature. This reflects the fact that the stacked classical
magnet represents a rather inefficient way of doing the bookkeeping
for the spin state. In the presence of only a few domain walls, it is
probably superior to keep track of the domain walls themselves. This
can be done in the framework of a continuous time algorithm,
recommended for future use, which is described for example in
Ref.~\onlinecite{riegercont}.

\section{Methods used and  
the fully frustrated three-leg ladder}
\label{sect:tll}

To introduce some of the concepts used repeatedly in this paper, let
us first consider the toy model of the fully frustrated three-leg
ladder depicted in Fig.~\ref{fig:fftll}. The interactions along the
rungs are ferromagnetic, as are those along the outer two legs. The
antiferromagnetic interactions along the middle leg make the ladder
fully frustrated.

The classical ground states are those states which minimise the number
of frustrated bonds. Since the bonds on the outer two legs belong to
only one plaquette, it is not favourable to frustrate these. By
contrast, frustrating one inner bond can put the two plaquettes it
belongs to into the ground state. One finds that there are three
sectors\cite{fn-dimtop} of ground states, and their Ising reversed
counterparts, depicted in Fig.~\ref{fig:fftll}. In all these states,
the top and bottom legs are ordered ferromagnetically. The staggered
sectors, with the spins on the top and bottom legs of opposite sign,
contain only one ground state each; by contrast, the columnar sector
has an extensive zero-temperature entropy per rung of ${\cal S} \equiv
\fracc{S_0}{(N k_B)}=G$, where $G=(\sqrt{5}+1)/2$\ is the golden mean,
which can be obtained by a transfer matrix approach.

Since each plaquette has exactly one frustrated bond, which is shared
with a neighbouring plaquette, one can represent the ground states by
placing dimers onto the frustrated bonds. This gives rise to a
hardcore dimer covering of the dual lattice, which in this case is a
two-leg ladder. Note that such a mapping is possible for all fully
frustrated models in which the elementary plaquettes are arranged to
share bonds. It does not, however, exist for all lattices; for
instance, if the plaquettes share sites rather than bonds, as is the
case for the kagome lattice, the hardcore nature of the dimer model is
lost. The advantage of this mapping is that, up to a global spin
reversal, there is a one-to-one correspondence between the ground
states and the hardcore dimer coverings. Therefore, restricting the
full Hilbert space to that of the dimer coverings yields a natural
implementation the projection onto the ground state. Below, we will
present a derivation of the transverse field Hamiltonian restricted to
the dimer (ground-state) manifold.

\subsection{The action of the transverse field -- mapping onto a 
quantum dimer Hamiltonian}
\label{subsect:tllqdm}

In the $S^z$\ basis, the transverse field operator $\Gamma S^x$\ is a
spin-flip operator, $(\fracc{\Gamma}{2})\left(\matrix{0&1\cr 1&0\cr}
\right)$. The leading-order quantum dynamics within the ground state
manifold is therefore that of single spin flips connecting different
ground states -- we need to consider multiple spin flips only in cases
where single spin flips are nowhere possible, as will be the case
in some of the examples discussed further down.

Flipping a spin without leaving the ground state manifold is possible
only if the spin is part of the same number of satisfied bonds as
frustrated ones, or, in other words, if it experiences zero net
exchange field. We refer to such spins as flippable spins. In dimer
language, such a spin is at the centre of a dimer plaquette of the
form 
$
\setlength{\unitlength}{3947sp}%
\begingroup\makeatletter\ifx\SetFigFont\undefined%
\gdef\SetFigFont#1#2#3#4#5{%
  \reset@font\fontsize{#1}{#2pt}%
  \fontfamily{#3}\fontseries{#4}\fontshape{#5}%
  \selectfont}%
\fi\endgroup%
\begin{picture}(154,155)(397,321)
\thicklines
\put(527,452){\circle{18}}
\put(528,344){\line( 0, 1){105}}
\put(527,345){\circle{18}}
\put(421,345){\circle{18}}
\put(421,344){\line( 0, 1){105}}
\put(421,452){\circle{18}}
\end{picture}
$\ or 
$ 
\setlength{\unitlength}{3947sp}%
\begingroup\makeatletter\ifx\SetFigFont\undefined%
\gdef\SetFigFont#1#2#3#4#5{%
  \reset@font\fontsize{#1}{#2pt}%
  \fontfamily{#3}\fontseries{#4}\fontshape{#5}%
  \selectfont}%
\fi\endgroup%
\begin{picture}(155,154)(533,319)
\thicklines
\put(664,343){\circle{18}}
\put(556,342){\line( 1, 0){105}}
\put(557,343){\circle{18}}
\put(557,449){\circle{18}}
\put(556,449){\line( 1, 0){105}}
\put(664,449){\circle{18}}
\end{picture}
$, and
flipping the spin exchanges the frustrated and the satisfied bonds and
thus corresponds to the elementary dimer move
$
\setlength{\unitlength}{3947sp}%
\begingroup\makeatletter\ifx\SetFigFont\undefined%
\gdef\SetFigFont#1#2#3#4#5{%
  \reset@font\fontsize{#1}{#2pt}%
  \fontfamily{#3}\fontseries{#4}\fontshape{#5}%
  \selectfont}%
\fi\endgroup%
\begin{picture}(154,155)(397,321)
\thicklines
\put(527,452){\circle{18}}
\put(528,344){\line( 0, 1){105}}
\put(527,345){\circle{18}}
\put(421,345){\circle{18}}
\put(421,344){\line( 0, 1){105}}
\put(421,452){\circle{18}}
\end{picture}
\leftrightarrow
\setlength{\unitlength}{3947sp}%
\begingroup\makeatletter\ifx\SetFigFont\undefined%
\gdef\SetFigFont#1#2#3#4#5{%
  \reset@font\fontsize{#1}{#2pt}%
  \fontfamily{#3}\fontseries{#4}\fontshape{#5}%
  \selectfont}%
\fi\endgroup%
\begin{picture}(155,154)(533,319)
\thicklines
\put(664,343){\circle{18}}
\put(556,342){\line( 1, 0){105}}
\put(557,343){\circle{18}}
\put(557,449){\circle{18}}
\put(556,449){\line( 1, 0){105}}
\put(664,449){\circle{18}}
\end{picture}
$.

The two staggered sectors of the ground state contain only the
two staggered dimer states.  The states in the columnar sector
can be obtained starting from the columnar dimer state by repeated
application of the elementary plaquette move
$
\setlength{\unitlength}{3947sp}%
\begingroup\makeatletter\ifx\SetFigFont\undefined%
\gdef\SetFigFont#1#2#3#4#5{%
  \reset@font\fontsize{#1}{#2pt}%
  \fontfamily{#3}\fontseries{#4}\fontshape{#5}%
  \selectfont}%
\fi\endgroup%
\begin{picture}(154,155)(397,321)
\thicklines
\put(527,452){\circle{18}}
\put(528,344){\line( 0, 1){105}}
\put(527,345){\circle{18}}
\put(421,345){\circle{18}}
\put(421,344){\line( 0, 1){105}}
\put(421,452){\circle{18}}
\end{picture}
\rightarrow
\setlength{\unitlength}{3947sp}%
\begingroup\makeatletter\ifx\SetFigFont\undefined%
\gdef\SetFigFont#1#2#3#4#5{%
  \reset@font\fontsize{#1}{#2pt}%
  \fontfamily{#3}\fontseries{#4}\fontshape{#5}%
  \selectfont}%
\fi\endgroup%
\begin{picture}(155,154)(533,319)
\thicklines
\put(664,343){\circle{18}}
\put(556,342){\line( 1, 0){105}}
\put(557,343){\circle{18}}
\put(557,449){\circle{18}}
\put(556,449){\line( 1, 0){105}}
\put(664,449){\circle{18}}
\end{picture}
$.

Therefore, in the staggered ground-state sectors the transverse field
has no effect since no elementary dimer moves are possible -- indeed,
there are no degenerate states which it can mix at $T=0$\ at any
finite order of the perturbation theory. This is different in the main
sector, where the transverse field lifts the macroscopic degeneracy
and promotes a particular linear combination of the classical ground
states to the true, quantum ground state.

In fact, it is now apparent that the transverse field Hamiltonian,
restricted to the classical ground state manifold, in dimer language
can be written as 
\bea H_{QDM}=-t\left
( |
\setlength{\unitlength}{3947sp}%
\begingroup\makeatletter\ifx\SetFigFont\undefined%
\gdef\SetFigFont#1#2#3#4#5{%
  \reset@font\fontsize{#1}{#2pt}%
  \fontfamily{#3}\fontseries{#4}\fontshape{#5}%
  \selectfont}%
\fi\endgroup%
\begin{picture}(155,154)(533,319)
\thicklines
\put(664,343){\circle{18}}
\put(556,342){\line( 1, 0){105}}
\put(557,343){\circle{18}}
\put(557,449){\circle{18}}
\put(556,449){\line( 1, 0){105}}
\put(664,449){\circle{18}}
\end{picture}
\rangle
\langle
\setlength{\unitlength}{3947sp}%
\begingroup\makeatletter\ifx\SetFigFont\undefined%
\gdef\SetFigFont#1#2#3#4#5{%
  \reset@font\fontsize{#1}{#2pt}%
  \fontfamily{#3}\fontseries{#4}\fontshape{#5}%
  \selectfont}%
\fi\endgroup%
\begin{picture}(154,155)(397,321)
\thicklines
\put(527,452){\circle{18}}
\put(528,344){\line( 0, 1){105}}
\put(527,345){\circle{18}}
\put(421,345){\circle{18}}
\put(421,344){\line( 0, 1){105}}
\put(421,452){\circle{18}}
\end{picture}
|+h.c.  \right) +v\left
( |
\setlength{\unitlength}{3947sp}%
\begingroup\makeatletter\ifx\SetFigFont\undefined%
\gdef\SetFigFont#1#2#3#4#5{%
  \reset@font\fontsize{#1}{#2pt}%
  \fontfamily{#3}\fontseries{#4}\fontshape{#5}%
  \selectfont}%
\fi\endgroup%
\begin{picture}(155,154)(533,319)
\thicklines
\put(664,343){\circle{18}}
\put(556,342){\line( 1, 0){105}}
\put(557,343){\circle{18}}
\put(557,449){\circle{18}}
\put(556,449){\line( 1, 0){105}}
\put(664,449){\circle{18}}
\end{picture}
\rangle \langle
\setlength{\unitlength}{3947sp}%
\begingroup\makeatletter\ifx\SetFigFont\undefined%
\gdef\SetFigFont#1#2#3#4#5{%
  \reset@font\fontsize{#1}{#2pt}%
  \fontfamily{#3}\fontseries{#4}\fontshape{#5}%
  \selectfont}%
\fi\endgroup%
\begin{picture}(155,154)(533,319)
\thicklines
\put(664,343){\circle{18}}
\put(556,342){\line( 1, 0){105}}
\put(557,343){\circle{18}}
\put(557,449){\circle{18}}
\put(556,449){\line( 1, 0){105}}
\put(664,449){\circle{18}}
\end{picture}
|+
|
\setlength{\unitlength}{3947sp}%
\begingroup\makeatletter\ifx\SetFigFont\undefined%
\gdef\SetFigFont#1#2#3#4#5{%
  \reset@font\fontsize{#1}{#2pt}%
  \fontfamily{#3}\fontseries{#4}\fontshape{#5}%
  \selectfont}%
\fi\endgroup%
\begin{picture}(154,155)(397,321)
\thicklines
\put(527,452){\circle{18}}
\put(528,344){\line( 0, 1){105}}
\put(527,345){\circle{18}}
\put(421,345){\circle{18}}
\put(421,344){\line( 0, 1){105}}
\put(421,452){\circle{18}}
\end{picture}
\rangle \langle
\setlength{\unitlength}{3947sp}%
\begingroup\makeatletter\ifx\SetFigFont\undefined%
\gdef\SetFigFont#1#2#3#4#5{%
  \reset@font\fontsize{#1}{#2pt}%
  \fontfamily{#3}\fontseries{#4}\fontshape{#5}%
  \selectfont}%
\fi\endgroup%
\begin{picture}(154,155)(397,321)
\thicklines
\put(527,452){\circle{18}}
\put(528,344){\line( 0, 1){105}}
\put(527,345){\circle{18}}
\put(421,345){\circle{18}}
\put(421,344){\line( 0, 1){105}}
\put(421,452){\circle{18}}
\end{picture}
|
\right),
\label{eq:qdmsquare}
\eea 
where the kinetic term
with $t=\Gamma/2$\ is generated by the transverse field. We have added a
diagonal term with coefficient $v$, which is zero for the transverse
field problem but which will be useful later on.

Note that this Hamiltonian has only nonpositive off-diagonal matrix
elements, so that the Perron-Frobenius theorem can be used to predict
a nodeless quantum ground state, which means that in it, the
amplitudes of all the configurations can be chosen to be real and
nonnegative.

Since a ground state and its Ising reversed counterpart both get
mapped onto the same dimer states, we have to show that the dimer
ground state is in fact the same as the transverse field ground
state.\cite{husedis} First we split the ground state manifold into two
submanifolds, namely those containing symmetric and antisymmetric
combinations of Ising-reversed pairs of states. The transverse
field Ising Hamiltonian does not connect these submanifolds, so that
it is block-diagonal.

All entries in the symmetric block continue to be of the same
(negative) sign so that the ground state in this block continues to be
nodeless. Since the full Hamiltonian is block-diagonal, the state
obtained by combining this ground state with a null state in the
antisymmetric block continues to be a nodeless eigenstate of the full
problem. Since there is only one nodeless eigenstate, the state
obtained by translating the dimer ground state into the spin ground
state is {\em the} ground state of the transverse field problem.

The entries in the antisymmetric sector have the same modulus as in
the symmetric one; however, they need not all have the same sign. If
we pick one spin state for each dimer state and collect those in the
up manifold, and their spin-reversed counterparts into the down
manifold, an entry will be negative when the transverse field connects
members of the up and down manifolds. The states can be sorted in a
way, e.g.\ by magnetisation, that the fraction of negative entries
vanishes in the thermodynamic limit. Whether the ground states in the
two sectors in the thermodynamic limit are degenerate then depends on
whether the wavefunction has substantial support on the states at zero
magnetisation, in which case they are not, or whether it is localised
away from them, in which case they are.  The example of the three-leg
ladder is special in that the up and down sectors can be chosen to be
entirely disconnected, so that an exact degeneracy trivially arises.

The connection between transverse field Ising model and quantum dimer
model we have established is useful for several reasons.  It affords
some insight into the structure of the problem we are studying in that
it provides a natural deformation of the transverse field Ising model
by switching on the potential term, i.e.\ by choosing a nonzero
$v$. The cases $|v|\gg t$\ are easily solved and can therefore provide
two anchors of the phase diagram containing the point we are
interested in. 

Moreover, there is the special point $v=t$, known as Rokshar-Kivelson
(RK) point, after the inventors of the model,\cite{Rokhsar88} where
the (nodeless) quantum ground state is an equal-amplitude
superposition of all classical ground states. Therefore, operators
which are diagonal in the dimer basis provide precisely the
expectation values of the corresponding {\em classical} operators. In
this spirit, switching on an infinitesimal transverse field implies
jumping the finite distance from $v=t$\ to $v=0$\ in the
Rokshar-Kivelson model (Fig.~\ref{fig:rkphase}), a manifestation
of the fact that the transverse field perturbation is
non-analytic. Deciding the ordering pattern of the transverse field
problem can therefore be accomplished if one can show that the
RK and the transverse field points are in the same
phase.

Moreover, the transverse field problem in itself provides a new
perspective on the Rokshar-Kivelson quantum dimer model, which was
proposed as a model of Anderson's (short-range) resonating valence
bond (RVB) physics.\cite{pwa87} It can be derived for a {\em
Heisenberg} antiferromagnet: the perturbative derivation uses the
nonzero overlap between different dimer configurations as expansion
parameter. This model is useful in a regime where the Heisenberg model
is in a phase dominated by valence bonds. The study of a transverse
field Ising model on a frustrated lattice can thus be used to gain
insight into the behaviour of Heisenberg magnets on the dual lattice.
This fact has been used by the present authors to identify a bona-fide
short-range RVB phase on the triangular lattice,\cite{MStrirvb} a goal
that had proven to be elusive on the square lattice, for which the
model was originally formulated.

Finally, we note parenthetically that for the mixed bond models, the
application of a longitudinal field is somewhat arbitrary as there is
a (gauge/Mattis) freedom of which bonds to call antiferromagnetic and
which ferromagnetic, as long as the odd condition is met. Thus any
state in the gauge invariant dimer representation can be (up to
topological restrictions) be represented by a maximally polarised
(long-range ordered) spin configuration, irrespective of the nature of
the dimer correlations. In the models where all bonds can be chosen to
be antiferromagnetic, this choice defines the natural gauge and thus
makes the application of a longitudinal magnetic field unambiguous.

\subsection{Maximally flippable states}
We now present a heuristic argument which is variational in spirit to
generate a candidate state for selection by the transverse field. This
state is the one which can gain the most energy from the transverse
field on account of being composed around a backbone configuration
having the strongest fluctuations possible. 

To start, we note that in order to gain energy from the transverse
field, spins have to have a component pointing in the $x$-direction,
which in the $S^z$-basis goes along with the component of the form
$\left[ |\uparrow\rangle + |\downarrow\rangle\right]/\sqrt{2}$. Those
states in which most of the spins can be flipped and hence put in a
superposition of up and down states thus stand to gain most energy
from the fluctuations induced by the transverse field. We therefore
identify the configuration with the largest number of flippable spins
(``maximally flippable state'') as the backbone of our candidate
ground state. The actual ground state naturally includes fluctuations
around the maximally flippable configuration, since it is these which
lead to the energy gain in the first place.\cite{coppersmith} The
energy gain due to alignment of the spins along the transverse field
in the quantum model is an {\em entropic} contribution to the free
energy in the classical, stacked model.

As an aside, we note that the classical correlations provide the
simplest first guess at the maximally flippable configurations and
hence at the ordering pattern. This is because configurations with
many flippable spins have a large number of neighbouring
configurations which differ only by a few spin flips. The correlations
they incorporate, even when not leading to long-range order, can thus
already be visible in the classical average where all ground states
are accorded equal weight.

For our three-leg ladder, the maximally flippable configuration is the
columnar one depicted in Fig.~\ref{fig:fftll}, since there, {\em all}
spins on the middle row (all of which point up) are flippable. We
therefore expect to find a state which has a ferromagnetic moment even
in the $z$-direction in addition to the polarisation in the
$x$-direction.

The selected state can incorporate structure in addition to that
apparent from the maximally flippable configuration. This follows from
the fact that although the backbone configuration maximises the number
of flippable spins, in fact not all the spins are {\em independently}
flippable. In the three-leg ladder, for example, flipping a spin on
the middle leg precludes flipping its neighbours, so that in effect
only half the spins are independently flippable. One can therefore
give two different recipes for constructing the quantum
state.

To contruct the first type of state, we take all the flippable spins
and polarise them in the $x$-direction disregarding the ground-state
constraint. Next, we reinstate the ground-state condition by
projecting out those components of the state which are not contained
in the classical ground-state manifold. We call this state the
uniform state since it treats all the flippable spins on the same
footing. 

The second type of state is obtained by identifying the largest set of
independently flippable spins and polarising those in the
$x$-direction. Since, starting from a maximally flippable
configuration, there are typically several choices for which set of
flippable spins to polarise, we call this state the hierarchical
state. It will have a lower symmetry than the maximally flippable
configuration unless all flippable spins are independently flippable.

The distinction between the uniform and the hierarchical states will
turn up in several times in this article, and it arises quite
naturally in other approaches. Note that this approach suggests yet
another class of alternative candidate configurations, namely those
which do not have the maximum number of flippable spins but
nonetheless maximise the number of spins which are flippable
independently.

The variational states thus obtained have the shortcoming that all
their component configurations are allocated equal weight. This is
clearly not optimal, since the ground state will in any case place
maximal weight on the individual, maximally flippable configurations.
Within a more elaborate variational framework, weights could be
accorded to the configurations depending on their number of flippable
spins.

In summary, the flippability approach suggested here identifies a
`saddle point' -- the maximally flippable configuration -- that will
be favoured due to the fluctuations around it, which are `softer' than
those around other configurations. We expect it to work as long as the
actual quantum wavefunction is concentrated on the maximally flippable
and nearby configurations. It will break down if the wavefunction has
the bulk of its support elsewhere, i.e.\ on configurations unrelated
to the maximally flippable one.

This point can be made more intuitively by considering the degenerate
perturbation theory as a hopping problem. Each classical ground state
configuration defines a point in the ground-state manifold. The
transverse field, by flipping spins, connects different
configurations, thus endowing the ground-state manifold with a graph
structure. The perturbation theory can be thought of as a hopping
problem on the graph thus defined. The large weight of maximally
flippable configurations follows from their high coordination, and the
magnetic ordering transition corresponds to a localisation transition
in the hopping problem.\cite{fn-hopping}

These ways of thinking are in close correspondence to the case of
thermal order by disorder.\cite{hfmrev} There, thermal fluctuations
({\sl out of} the ground state manifold) provide a large entropic
weighting to the states allowing the softest fluctuations.  These
states are then selected as $T\rightarrow 0^+$\ provided their
enhanced weight is not swamped by the combined fluctuational and
configurational entropy of the less favoured states.\cite{pyroshlo}
However, when the thermal fluctuations increase in strength with
increasing temperature, they destroy the order which they were
instrumental in establishing in the first place, as happens for
$\Gamma$\ large in our problem, as discussed in the following
paragraphs.

\subsection{The opposite limit: Landau-Ginzburg analysis 
for $\mathbf{\Gamma \gg J}$} In addition to doing the (hard)
degenerate perturbation theory for small transverse fields, we can use
an alternative approach for determining the state of the quantum
magnet which is made tractable by virtue of the simplicity of the
transverse-field term. Consider the problem where the relative sizes
of exchange and transverse field are inverted, namely where $\Gamma
\gg J$. In the limit $J/\Gamma=0$, the ground state is a simple
paramagnet in the $S^z$\ basis: all spins are polarised along the
$+x$-direction explicitly selected by the field. In addition, it is
gapped: the lowest excitations are spin flips which each cost an
energy of $\Gamma$; this makes it possible to perturb about this state
by switching on a weak exchange. This contrasts to the case where the
transverse field is replaced by an XY exchange. Here, the large
$J_{XY}$\ problem is not exactly soluble, and so it is not possible to
perturb about it -- indeed, it may even be gapless.

As the perturbing $J$\ is switched on in addition to the transverse
field, the excitations acquire a dispersion, typically -- but not
always -- already to first order in $J/\Gamma$. The dispersion to
first order is simply given by the Fourier transform of the
interaction matrix of the lattice.  
For sufficiently large
$J$, there can be an ordering transition, which corresponds to a
macroscopic occupation of the softest mode(s).

To generate the state to which the leading transition takes place, one
has to combine knowledge of the soft modes with lattice
symmetry considerations to construct a Landau-Ginzburg-Wilson
action. This program is carried out in detail for the fully-frustrated
hexagonal magnet below and follows the work on layered magnets by
Blankschtein and coworkers.\cite{Blankschtein84,blankvillain}

The combinations of the soft modes dictated by symmetry considerations
yield the ordering pattern which is established as $J/\Gamma$\ is
increased. In addition, by also analysing the Landau-Ginzburg-Wilson
action with standard methods of the renormalisation group, one can
obtain information on the nature of the transition into the ordered
configuration, and it turns out that these transitions are generally
not Ising transitions, as the most naive guess would suggest.  By
analysing the potential presence of dangerously irrelevant terms in
the action, one can even guess at further symmetry breaking.

This analysis of course has the usual limits associated with
mean-field theories. Most important in this context is the possibility
of further phase transitions out of the ordered phase before we reach
the regime of infinitesimal transverse fields, where the `small'
parameter $J/\Gamma\rightarrow\infty$. Another scenario is the absence
of any phase transition, so that the magnet remains disordered at all
couplings. Although the excitation dispersion may soften at particular
points in the Brioullin zone, fluctuations may be sufficiently strong
to prevent ordering at any coupling. However, even in this situation,
the large $\Gamma$\ approach may be used successfully if the expansion
in powers of $J/\Gamma$\ is carried to sufficiently high order and
combined with a non-perturbative analysis such as that given by the
use of Pade approximants. In Ref.~\onlinecite{donmar}, this program
has been carried out for the sawtooth chain (see
Fig.~\ref{fig:alllattices}).

It is worth pointing out that this approach presents, ultimately, a
soft-spin analysis in that the size of the ordered moment can vary
from site to site so that, as the mode amplitude increases,
non-linearities become important. However, even for our hard Ising
spins, a difference in the size of the ordered moment does have a
meaning. As an illustration, consider the uniform state of the
three-leg ladder, defined above. Here, every spin on the middle row is
fluctuating and thus has $\left|\left<S_z\right>\right|<1/2$, whereas
the remaining spins have $\left<S_z\right>$=1/2, being fully polarised
along the $z$-axis. This difference is indeed found in the large
$\Gamma$\ analysis and can thus be interpreted as being due to the
fluctuations induced by the transverse field.

In detail, for the three-leg ladder, the Fourier transform of the 
interaction matrix is 
(labeling the sites on the rung 1 to 3 from top to bottom)
\beq
\frac{J}{2}  \left(\matrix{ 2\,\cos (k) & 
1 & 0 \cr 1 & -2\,\cos (k) & 1 \cr 0 & 1 & 2\,
    \cos (k) \cr  } \right)\ ,
\eeq
from which we obtain a degeneracy in the
dispersion relation. One minimal mode, with wavevector $q=0$, has
eigenvector $(1,-2+\sqrt{6},1)$; the other has $q=\pi$\ and
eigenvector $(1,-2-\sqrt{6},1)$. Here, $q$\ is the wavevector along
the ladder, and the entries in the eigenvector denote the amplitudes
of the top, middle and bottom sites ($\sqrt{6}\simeq2.45$).

This degeneracy is accidental in that the states are not related by
any symmetry operations, and we have not encountered this effect in
any of the more regular lattices we are studying. Quite generally,
however, in lattices with sites of different coordination, the
soft-mode analysis might suggest states which fare very poorly under
the hard-spin constraint, as is the case for the $q=\pi$\ mode in this
example (see below). A more appropriate ordering pattern may
nonetheless show up as an alternative, possibly only {local} minimum
in the dispersion relation.

The $q=0$\ mode can easily be identified as the actual ordering
pattern (see below, exact diagonalisation), with a moment on the
middle row reduced by the fluctuations. One reason the competing state
loses out eventually is because it has a reduced moment on the sites
which are in fact not allowed to fluctuate in the small-$\Gamma$\
limit: there, the leading term in the perturbation theory flips spins
with an equal number of frustrated and satisfied bonds, and the
reduced moment sites in this state have odd coordination and are thus
never flippable.

\subsection{Exact diagonalisation}

For the three-leg ladder, one can attack the transverse field problem
by exact diagonalisation of the degenerate perturbation
theory. Systems with up to fourteen rungs are easily accessible
numerically. The results are depicted in Fig.~\ref{fig:tllplot}.
We do indeed find a state with ferromagnetic order along
the centre row, as predicted by the flippability analysis. The
selected state is the uniform columnar one, as the gap extrapolates to
a finite value for large system sizes and hence the breaking of
translational symmetry accompanying the hierarchical state is absent.

The infinitesimal transverse field generates a nonzero polarisation in
the $x$-direction, corresponding to an ground-state energy per spin of
$E_0=-(0.302\pm0.001)|\Gamma|$\ in excess of the classical value. The
finite moment $\langle S_z\rangle$\ already present in the classical
ground state average (where it equals $1/(2(2G-1))\simeq0.224$), is
discontinuously enhanced as $\Gamma$\ is switched on.  As the system
size is increased, the maximally flippable state and its neighbours
rapidly gain in weight, whereas at the other extreme, those states
without flippable spins have zero amplitude in the ground state
wavefunction.

These results tie in with the picture provided by the mapping onto a
quantum dimer model. The selected columnar dimer phase, which extends
all the way to $v=-\infty$, terminates at the RK point,
beyond which it gives way to the staggered dimer phase, as depicted in
Fig.~\ref{fig:rkphase}.\cite{fn-nosym}

\subsection{Relation to XY perturbations}

Whereas the transverse field induces single spin flips, an XY exchange
flips neigbouring pairs of antialigned spins. For lattices with even
coordination and all bonds antiferromagnetic, the XY flips are a
subset of the two spin flips generated by two successive operations of
the transverse field. 

For lattices with odd coordination, however, the transverse field can
never flip a spin without leaving the ground state manifold, so that
the XY moves can show up as the {\em leading} terms, in order
$\Gamma^2/J$, in the degenerate perturbation theory. These matrix
elements are negative, corresponding to a ferromagnetic XY
exchange.  This will be of importance for the fully frustrated
hexagonal lattice. Moreover, in the case of the pentagonal lattice,
where inequivalent sites with even and odd coordination exist, the
moves generated by an XY perturbation are entirely distinct from those
generated by the transverse field.

\section{The pentagonal lattice}
\label{sect:pent}

The pentagonal lattice, depicted in Fig.~\ref{fig:pent}, tiles the
plane with (irregular) pentagons. It can be obtained from the
hexagonal lattice by cutting each hexagon in half with a set of
parallel lines (`cuts'). Classically, the Ising antiferromagnet on
this lattice is disordered and has a finite ground-state entropy per
spin of ${\cal{S}}/k_B=0.234$.\cite{pent} The ground states can be
represented by a hardcore dimer model as described above.

The flippability analysis for this magnet is straightforward. To
lowest order in $\Gamma$, only the spins along the cuts can be flipped
since only they have even coordination. Therefore, the spins along the
sawtooths follow an antiferromagnetic pattern, since the maximally
flippable configuration (depicted in Fig.~\ref{fig:pent}) has all
frustrated bonds associated with spins on the cuts. This configuration
is closely related to the fully frustrated three-leg ladder in that
the spins along the cuts are effectively decoupled from the other
spins in the system and the transverse field generates the same
Hamiltonian as for the ladder. Thence, the ordering pattern along the
cuts will be the $q=0$\ pattern described above, with a 
$q=\pi$\ modulation transverse to the cuts.

For completeness, we now perform the soft-mode analysis for this
problem. The interaction matrix is

\bea \frac{J}{2}\left(\matrix{ 2\,\cos q_x & 1 & {e^{-i\,q_y}} \cr 1
& 0 & 1 + {e^{-i\,q_x}} \cr {e^{i\,q_y}} & 1 + {e^{i\,q_x}} & 0 \cr }
\right) \nonumber \eea

There is a line where the eigenvalues attain the global minimum
$-J(1+\sqrt{3})/2$\ at wavectors $(\pi,y)$. The eigenvectors $\left(
({-1 - {\sqrt{3}}){{e^{-i\,q_y}}}},{e^{-i\,q_y}},1\right) $\
correspond to states with frustrated bonds on the sawtooths. As
mentioned above, these do not lead to flippable spin arrangements to
lowest order in the degenerate perturbation theory in $\Gamma$, and
this state is therefore not competitive in the small $\Gamma$\
limit. There is, in addition, one further local minimum
$-J\sqrt{6}/2$\ at wavevector $(0,\pi)$. The eigenvector $\left( -2 +
{\sqrt{6}},-1,1\right)$\ corresponds to the maximally flippable state
(and agrees with the one given by the three-leg ladder). We therefore
expect a phase diagram with at least three transitions coming from
large $\Gamma$, namely first from the disordered into the `sawtooth
state' followed by a transition into the maximally flippable state,
which finally terminates in the disordered classical phase at
$\Gamma=0$.  There may be a separate transition which selects one or a
combination of the sawtooth states -- we have not investigated
this. The complicated structure of this phase diagram is of course to
a large degree a consequence of the presence of inequivalent sites.

\section{The triangular antiferromagnet}
\label{sect:triang}

The two-dimensional Ising antiferromagnet on the triangular lattice
has power-law spin correlations and an extensive entropy (${\cal
S}/k_B= 0.323$) at $T=0$.\cite{wannier,houtappel,stephensontri} We
first describe a number of useful mappings and then discuss their
implications.

\subsection{Mappings to height and dimer models}

In Fig.~\ref{fig:triooo}, we have displayed a particular spin
configuration. The corresonding dimer configuration (also shown) is
constructed as in Sect.~\ref{sect:tll}). The effect of the transverse
field can easily be stated in terms of the dimer mapping in a more
general setting than the one discussed in Sect.~\ref{sect:tll}).  If
we choose a closed loop of bonds on the dual lattice which are
alternately empty and occupied by a dimer, and exchange the occupied
with the empty bonds, this corresponds to flipping all the spins
inside the closed loop. The leading order effect of the transverse
field is therefore to produce such a rearrangement for the loop
enclosing the smallest number of sites, and the relevant order in
perturbation theory is given by the number of enclosed spins.  The
loop in question is typically the shortest closed loop of even length
on the dual lattice. In this case, this is a loop of length six
enclosing one site (see Fig.~\ref{fig:triooo}), but we will encounter
the case of more than one enclosed site later on.

Next, we turn our attention to the mapping to a height model.  The
height configuration is given by the numbers in
Fig.~\ref{fig:triooo}. The height variables are defined on the sites
of the lattice.\cite{Blote82} Height differences are determined as
follows. If one crosses a (no) dimer when going along a bond in a
clockwise direction around a triangle pointing up, the height changes
by 2 ($-1$). If one goes the opposite direction, the height change is
$-2$ (+1). One can easily check that this prescription has the
property that the height change around the three sides of any
triangle is zero. Moreover, any closed path can be decomposed into a
sum of paths around individual triangles, and thus the height change
between any two sites is indepent of the path chosen to go between
them.

The existence of the height model has important ramifications for the
classical model. Under a set of reasonable assumptions,\cite{Blote82}
the result follows that the ground state correlations are
algebraic. Combined with our criterion for quantum ordering
(Sect.~\ref{subsect:criterion}), one expects all magnets allowing a
height mapping to order. The ordering correlations are then determined
by the configuration which is as flat (in height space) as possible.

This raises the question under which conditions a height mapping
exists. A set of sufficient conditions is that, firstly, the lattice
be composed of bond sharing frustrated units (allowing a dimer model)
and secondly, its dual lattice be bipartite. This is required for
giving a set of consistent rules for computing height differences. In
the cases we encounter below where these conditions are not met, we
will see that a height mapping is absent, consistent with their
disordered classical correlations.

\subsection{Ordering behaviour of the triangular IAFM} 
We first study this problem for small transverse field with $h=0$.
The maximally flippable configuration is the one depicted in
Fig.~\ref{fig:triooo}. It has a unit cell of three sites (two spins
pointing up and one down). This is also the flat (height)
configuration, as well as being the state encountered at $v=-\infty$
in the appropriate quantum dimer model:\cite{fn-Orland,henabs} 
\bea
H_{ QDM} = -t\left( | \setlength{\unitlength}{3158sp}%
\begingroup\makeatletter\ifx\SetFigFont\undefined%
\gdef\SetFigFont#1#2#3#4#5{%
  \reset@font\fontsize{#1}{#2pt}%
  \fontfamily{#3}\fontseries{#4}\fontshape{#5}%
  \selectfont}%
\fi\endgroup%
\begin{picture}(194,185)(318,260)
\thicklines \put(468,421){\circle{18}}
\put(358,420){\line( 1, 0){108}}
\put(342,377){\circle{18}} \put(360,421){\circle{18}}
\put(396,284){\circle{18}}
\multiput(434,282)(5.48824,9.14707){11}{\makebox(8.3333,12.5000){\SetFigFont{7}{8.4}{\rmdefault}{\mddefault}{\updefault}.}}
\multiput(340,378)(5.57647,-9.29412){11}{\makebox(8.3333,12.5000){\SetFigFont{7}{8.4}{\rmdefault}{\mddefault}{\updefault}.}}
\put(434,284){\circle{18}}
\put(488,377){\circle{18}}
\end{picture}
 \rangle
\langle
\setlength{\unitlength}{3158sp}%
\begingroup\makeatletter\ifx\SetFigFont\undefined%
\gdef\SetFigFont#1#2#3#4#5{%
  \reset@font\fontsize{#1}{#2pt}%
  \fontfamily{#3}\fontseries{#4}\fontshape{#5}%
  \selectfont}%
\fi\endgroup%
\begin{picture}(194,185)(318,93)
\thicklines \put(468,117){\circle{18}}
\put(358,118){\line( 1, 0){108}}
\put(342,161){\circle{18}} \put(360,117){\circle{18}}
\put(396,254){\circle{18}}
\multiput(434,256)(5.48824,-9.14707){11}{\makebox(8.3333,12.5000){\SetFigFont{7}{8.4}{\rmdefault}{\mddefault}{\updefault}.}}
\multiput(340,160)(5.57647,9.29412){11}{\makebox(8.3333,12.5000){\SetFigFont{7}{8.4}{\rmdefault}{\mddefault}{\updefault}.}}
\put(434,254){\circle{18}}
\put(488,161){\circle{18}}
\end{picture}
|+h.c.  \right) +v\left
( |
\setlength{\unitlength}{3158sp}%
\begingroup\makeatletter\ifx\SetFigFont\undefined%
\gdef\SetFigFont#1#2#3#4#5{%
  \reset@font\fontsize{#1}{#2pt}%
  \fontfamily{#3}\fontseries{#4}\fontshape{#5}%
  \selectfont}%
\fi\endgroup%
\begin{picture}(194,185)(318,260)
\thicklines \put(468,421){\circle{18}}
\put(358,420){\line( 1, 0){108}}
\put(342,377){\circle{18}} \put(360,421){\circle{18}}
\put(396,284){\circle{18}}
\multiput(434,282)(5.48824,9.14707){11}{\makebox(8.3333,12.5000){\SetFigFont{7}{8.4}{\rmdefault}{\mddefault}{\updefault}.}}
\multiput(340,378)(5.57647,-9.29412){11}{\makebox(8.3333,12.5000){\SetFigFont{7}{8.4}{\rmdefault}{\mddefault}{\updefault}.}}
\put(434,284){\circle{18}}
\put(488,377){\circle{18}}
\end{picture}
\rangle \langle
\setlength{\unitlength}{3158sp}%
\begingroup\makeatletter\ifx\SetFigFont\undefined%
\gdef\SetFigFont#1#2#3#4#5{%
  \reset@font\fontsize{#1}{#2pt}%
  \fontfamily{#3}\fontseries{#4}\fontshape{#5}%
  \selectfont}%
\fi\endgroup%
\begin{picture}(194,185)(318,260)
\thicklines \put(468,421){\circle{18}}
\put(358,420){\line( 1, 0){108}}
\put(342,377){\circle{18}} \put(360,421){\circle{18}}
\put(396,284){\circle{18}}
\multiput(434,282)(5.48824,9.14707){11}{\makebox(8.3333,12.5000){\SetFigFont{7}{8.4}{\rmdefault}{\mddefault}{\updefault}.}}
\multiput(340,378)(5.57647,-9.29412){11}{\makebox(8.3333,12.5000){\SetFigFont{7}{8.4}{\rmdefault}{\mddefault}{\updefault}.}}
\put(434,284){\circle{18}}
\put(488,377){\circle{18}}
\end{picture}
|+
|
\setlength{\unitlength}{3158sp}%
\begingroup\makeatletter\ifx\SetFigFont\undefined%
\gdef\SetFigFont#1#2#3#4#5{%
  \reset@font\fontsize{#1}{#2pt}%
  \fontfamily{#3}\fontseries{#4}\fontshape{#5}%
  \selectfont}%
\fi\endgroup%
\begin{picture}(194,185)(318,93)
\thicklines \put(468,117){\circle{18}}
\put(358,118){\line( 1, 0){108}}
\put(342,161){\circle{18}} \put(360,117){\circle{18}}
\put(396,254){\circle{18}}
\multiput(434,256)(5.48824,-9.14707){11}{\makebox(8.3333,12.5000){\SetFigFont{7}{8.4}{\rmdefault}{\mddefault}{\updefault}.}}
\multiput(340,160)(5.57647,9.29412){11}{\makebox(8.3333,12.5000){\SetFigFont{7}{8.4}{\rmdefault}{\mddefault}{\updefault}.}}
\put(434,254){\circle{18}}
\put(488,161){\circle{18}}
\end{picture}
\rangle \langle
\setlength{\unitlength}{3158sp}%
\begingroup\makeatletter\ifx\SetFigFont\undefined%
\gdef\SetFigFont#1#2#3#4#5{%
  \reset@font\fontsize{#1}{#2pt}%
  \fontfamily{#3}\fontseries{#4}\fontshape{#5}%
  \selectfont}%
\fi\endgroup%
\begin{picture}(194,185)(318,93)
\thicklines \put(468,117){\circle{18}}
\put(358,118){\line( 1, 0){108}}
\put(342,161){\circle{18}} \put(360,117){\circle{18}}
\put(396,254){\circle{18}}
\multiput(434,256)(5.48824,-9.14707){11}{\makebox(8.3333,12.5000){\SetFigFont{7}{8.4}{\rmdefault}{\mddefault}{\updefault}.}}
\multiput(340,160)(5.57647,9.29412){11}{\makebox(8.3333,12.5000){\SetFigFont{7}{8.4}{\rmdefault}{\mddefault}{\updefault}.}}
\put(434,254){\circle{18}}
\put(488,161){\circle{18}}
\end{picture}
|\right) \ .
\label{eq:triqdm}
\eea 

In fact, this state has a net magnetisation since its three-sublattice
strucutre is $(1,1,-1)$, with all spins pointing up being
flippable. The uniform state will thus have a sublattice structure of
the form $(a,a,-b)$, with $b>a$, whereas the hierarchical one will
have $(c,0,-c)$.

All the evidence therefore points towards a three sublattice quantum
ordering pattern, with details to be determined.  Such correlations
were previously found by a Landau-Ginzburg-Wilson (LGW) analysis of
ferromagnetically-stacked triangular lattices,\cite{Blankschtein84}
applicable to our problem by virtue of the mapping presented in
Sect.~\ref{subsect:mapstack}.  This LGW analysis finds an action which
is XY symmetric up to sixth order, where an XY symmetry breaking,
six-fold clock term appears.

The resulting phase diagram in the temperature-field ($T-\Gamma$)
plane, depicted in Fig.~\ref{fig:jose}, is quite remarkable.  At
$T=0$, the triangular TFIM undergoes a quantum phase transition which
is in the 3-d XY universality class,\cite{Blankschtein84} where the
clock term is dangerously irrelevant, so that the transition is into a
phase with clock-symmetry breaking as well. 

At finite temperature, however, the transverse field Ising model maps
onto a classical stacked magnet of {\em finite size}, $L^\tau$, in the
imaginary time direction, so that its behaviour crosses over to being
two-dimensional as the correlation length becomes comparable to
$L^\tau$; in this regime, we thus have to consider the properties of a
two-dimensional XY model with a clock term. Here one finds\cite{jose}
two finite-temperature Kosterlitz-Thouless transitions bordering an
extended critical phase. The phase diagram in Fig.~\ref{fig:jose} is
reliable in the region near the three dimensional XY transition, where
the order parameters are small.
 
Depending on the sign of the six-fold clock term, the quantum ordered
state is predicted to have the sublattice structure of either the
uniform or the hierarchical state. In order to confirm the predicted
ordering pattern and to find out the precise sublattice structure, we
have carried out quantum Monte Carlo simulations on the triangular
TFIM. The results are depicted in Figs.~\ref{fig:tribragg} and
\ref{fig:triagnew}.  A Bragg peak is clearly visible at $q_x=2\pi/3$,
as expected. The sublattice structure of the form $(1,0,-1)$\ is also
clearly borne out, showing that the hierarchical state without a net
moment is selected. Since the quantum dimer state at $v=-\infty$\ is
the uniform, magnetised state, this implies a transition for
$v<0$. Preliminary studies by us indicate that such a transition is
indeed present close to $v=0$. Finally, we note that a finite
longitudinal field ($|h| < 6J$) on its own selects the same uniform
state.

\section{Villain's odd model: the fully frustrated square lattice}
\label{sect:villain}

A great deal is known about the square FFIM, especially when we
combine the knowledge about the different models it is equivalent to
by virtue of the mappings presented above.  It turns out to shadow the
triangular IAFM in practically all important qualitative respects.
Its classical ground-state entropy ${\cal
S}=C/\pi\simeq0.241$,\cite{mefvillain} where $C$\ is Catalan's
constant. Since it consists of bond-sharing squares with a bipartite
dual lattice, it is critical at $T=0$. The corresponding dimer model
Hamiltonian is given by Eq.~\ref{eq:qdmsquare} and has a phase diagram
like that pictured in Fig.~\ref{fig:rkphase}. The columnar phase
corresponds to the maximally flippable state which, in dimer language,
is obtained by covering the square lattice with infinite dimer ladders
with the configuration
%of the type 
depicted on the right panel of Fig.~\ref{fig:fftll}. An LGW
analysis\cite{blankvillain} again finds an XY action with an XY
symmetry breaking clock anisotropy at higher order, which, depending
on its sign, selects either the uniform or the hierarchical (also
known as plaquette) state.

This makes it clear that there is quantum ordering in the FFIM on the
square lattice into a state with translational symmetry breaking. We
have not studied this problem numerically ourselves but we refer the
reader to the literature on the ferromagnetically stacked
magnet\cite{blankvillain,sachjala} and the quantum dimer
model.\cite{orlandsquare,leung}

\section{Pyrochlore in two dimensions: the square 
lattice with crossings}
\label{sect:twodpyro}

This lattice, also known as the checkerboard or two-dimensional
pyrochlore lattice, is made up of a square lattice with nearest
neighbour interactions which has, in addition, crossing next-nearest
neighbour interactions on alternate plaquettes as depicted in
Fig.~\ref{fig:sqcrossheight}. The plaquettes with crossings are
equivalent to tetrahedra in that they contain four sites all
interacting equally with one another. Since these tetrahedra are
arranged to share sites, as is the case in the pyrochlore lattice in
three dimensions, from which it can in fact be obtained by a
projection in a $\left<111\right>$-direction; for a picture, see
Ref.~\onlinecite{pyroshlo}b. 

The ground-state condition for the classical Ising model is that each
tetrahedron have zero magnetisation. There are six such states for
each tetrahedron, each with two spins up and two down.  The Ising
model on the pyrochlore lattice is equivalent to the ice
model,\cite{andersonpyro} the ground state entropy of which was
calculated exactly in two dimensions by Lieb:\cite{liebice}
${\cal{S}}/k_B=\frac{3}{4}\ln\frac{4}{3}\approx0.216$. 
This model can be mapped onto a six-vertex model
(each vertex encoding one of the single-tetrahedron ground states),
and hence onto a non-intersecting loop model, which in two dimensions
guarantees the existence of a height model. As discussed above, this
quite generally implies classically critical ground-state
correlations, which have indeed been found.\cite{Liebmann86}

Here, we present a short derivation of the height model which we then
use to determine the quantum ordering behaviour; refer to
Fig.~\ref{fig:sqcrossheight}. We define a set of one-dimensional
heights which reside on the vertices of the square lattice dual to the
plaquettes without crossing interactions.  Next, we assign an
orientation (clockwise or anticlockwise) to each tetrahedron, so that
neighbouring tetrahedra have opposite orientations. Since the lattice
dual to the tetrahedra is bipartite, this can be done
consistently. The rule for the height differences is as follows. If
going from one site of the height lattice to another one passes over
an up (down) spin, one increases (decreases) the height by one
provided the spin was crossed in the direction given by the
orientation of the tetrahedra it belongs to. In the opposite
direction, one decreases (increases) the height by one. This generates
a consistent assignment of the heights since going around a unit cell
of the height lattice generates zero height difference by virtue of
the ground-state two-up two-down condition, and because each closed
path on the height lattice can be decomposed into a combination of
such elementary loops.\cite{newbarkema}

We next consider the action of a longitudianl field of strength $h$\
in the absence of a transverse field. For $h<2J$, all ground states
remain degenerate since they have zero net moment.  At $h=2J$, the
applied field is strong enough to surmount the exchange field and it
generates a spin-flop transition to a manifold of states with three
spins up and one down in each tetrahedron. These states continue to
have an extensive entropy but one which is reduced compared to the
low-field value. In addition, the classical correlations in this
regime, to which we allocate the name IM (`intermediate'-field
regime), continue to be critical. This result follows from another
mapping of those states onto a dimer model which in turn generates a
height model.

This mapping is obtained as follows. Consider a dimer model on the
square lattice dual to the tetrahedra. For each spin pointing down,
place a dimer, centred on this spin, with its ends located at the
points of the dual lattice denoting the centres of the tetrahedra the
down spin belongs to. Since each tetrahedron has exactly one down
spin, and since each spin is shared by two tetrahedra, each classical
IM ground state generates a hardcore dimer covering of the (dual)
square lattice and vice versa. Such a square-lattice dimer model can
be mapped onto a height model, from which the criticality of the
correlations in this regime follow. In the IM-phase, the entropy per
spin is a quarter of the value found in the Villain model, as now
there are four spins per dimer rather than one: ${\cal S}/k_B=0.073$.

We emphasize that the field leaves the entropy unchanged for a finite
range of $h$\ and then reduces to a lower value without eliminating it
completely, while inducing a transition between two critical states;
this again persists over a finite range of fields. At $h=6J$, there
finally occurs a transition to the fully polarised state.

Next, consider tilting the field such that $\Gamma\ll h$; this enables
us to generate a perturbation theory controlled by the small parameter
$\Gamma/h$.  The transverse field induces matrix elements between the
states corresponding to the classical ground state
configurations. Both in the low-field and the IM regime, the
degeneracy is not lifted until fourth order in perturbation
theory. The reason is that connecting two ground states requires
flipping a closed loop of alternating spins which passes through an
even number of sites of any tetrahedron. The shortest such loop has
length four and is thus generated at fourth order in perturbation
theory. The lower-order terms induce only a diagonal shift in the
energies but this shift is the same for all states.
%as explained in
%analogousSect.~\ref{sect:kagome}.

The short flippable loop, depicted in Fig.~\ref{fig:sqcrossheight},
has different interpretations in the two phases.  For the low-field
phase, it corresponds to changing the height of a plaquette whose four
neighbouring heights are equal. In a manner analogous to the
ground-state selection on the triangular lattice, this leads to the
selection of the flat state in height language. In spin language, the
flat state is a Neel state on the square lattice underlying the
square-lattice with crossings.

In the IM phase, the quadruple spin-flip again leads to the RK model
at $v=0$\ and $t\propto \Gamma^4/J^3$ (see
Eq.~\ref{eq:qdmsquare}). This move connects different ground states
since the total magnetisation of each tetrahedron remains unchanged
and the effect of flipping the four spins is to generate the familiar
dimer plaquette move
$
\setlength{\unitlength}{3947sp}%
\begingroup\makeatletter\ifx\SetFigFont\undefined%
\gdef\SetFigFont#1#2#3#4#5{%
  \reset@font\fontsize{#1}{#2pt}%
  \fontfamily{#3}\fontseries{#4}\fontshape{#5}%
  \selectfont}%
\fi\endgroup%
\begin{picture}(154,155)(397,321)
\thicklines
\put(527,452){\circle{18}}
\put(528,344){\line( 0, 1){105}}
\put(527,345){\circle{18}}
\put(421,345){\circle{18}}
\put(421,344){\line( 0, 1){105}}
\put(421,452){\circle{18}}
\end{picture}
\leftrightarrow
\setlength{\unitlength}{3947sp}%
\begingroup\makeatletter\ifx\SetFigFont\undefined%
\gdef\SetFigFont#1#2#3#4#5{%
  \reset@font\fontsize{#1}{#2pt}%
  \fontfamily{#3}\fontseries{#4}\fontshape{#5}%
  \selectfont}%
\fi\endgroup%
\begin{picture}(155,154)(533,319)
\thicklines
\put(664,343){\circle{18}}
\put(556,342){\line( 1, 0){105}}
\put(557,343){\circle{18}}
\put(557,449){\circle{18}}
\put(556,449){\line( 1, 0){105}}
\put(664,449){\circle{18}}
\end{picture}
$.

Carrying over the results from the fully frustrated square lattice, we
expect the system to order into a flat (height) phase which
corresponds to a columnar dimer phase. In spin language, this phase
differs from the Neel low-field phase, from which it can be obtained
by flipping half of the down spins, e.g. those which are located on
the top-left hand corner of one sublattice of tetrahedra.

The complete phase diagram for this magnet in the $h-\Gamma$\ plane
can be constructed from this.  The simplest phase diagram incorpoating
our results is displayed in Fig.~\ref{fig:sqcrossphase}.

\section{The kagome lattice}
\label{sect:kagome}

In the kagome Ising antiferromagnet, depicted in
Fig.~\ref{fig:alllattices}, the nearest-neighbor exchange couplings
are uniform and antiferromagnetic.  The ground-state entropy is
finite,\cite{Kano53} ${\cal S}_{kag} = 0.502$, and is more than half
of the maximum paramagnetic value (${\cal S}_{para} = \ln 2$) in
contrast to the triangular case (${\cal S}_{tri} = 0.323$).
Furthermore ${\cal S}_{kag}$ is close in value to that obtained by the
Pauling approximation,\cite{Liebmann86,Pauling38} ${\cal S}_{Pauling}
= \ln 2 + \frac{2}{3} \ln \frac{3}{4} \approx 0.501$, where the
triangles are considered independently; this suggests that
spin-spin correlations in the kagome IAFM are extremely weak.  This is
indeed the case, and the model remains classically disordered at all
temperatures.\cite{Kano53}

An important feature or the kagome lattice is that the frustrated
units (triangles) are arranged to share sites rather than bonds.  This
precludes the mapping of the kagome ground states to a hardcore dimer
model on the dual lattice (known as the diced lattice); this also
precludes a mapping to a height model. It is thus an excellent
candidate for a disordered quantum magnet.
The physics of this model has been discussed in moderate detail in
Ref.~\onlinecite{mcs2000}; here, we fill in the missing detail,
repeating some material telegraphically for coherence and
convenience.

Application of a longitudinal field, $|h| < 4J$, to this magnet leads
to the development of a ferromagnetic moment coexisting with a reduced
but extensive entropy and {\em critical} spin correlations.  The
result follows in a way formally similar to the $\Gamma=0$\ phases in
Sect.~\ref{sect:twodpyro}. Each triangle has two up and one down spin,
and denoting each down spin by a dimer with endpoints in the centres
of the triangles that share it, we obtain a bijective mapping between
the ground states and the hardcore dimer covering of the hexagonal
lattice; see Fig.~\ref{fig:kagphase}. This implies, by virtue of the
associated height mapping, critical correlations. The entropy per
spin, while still nonzero, is reduced to the value of ${\cal S}={\cal
S}_{tri}/3=0.108$; the difference to the triangular lattice arises
from the fact that the number of spins per dimer is different on
account of the inequivalent mappings used to arrive at the dimer
model.

We emphasize that this result is rather unusual. Starting from a
disordered magnet, we obtain a {\em critical} state with nonzero
moment and extensive entropy upon application of an infinitesimal
field; these properties, including the criticality, persist for a
finite range of field strengths. 

Next, consider applying a small transverse field, $\Gamma \ll |h|$, in
addition to the longitudinal one ($0 < |h| < 4J$). The resulting
perturbation theory, controlled by the parameter $|\Gamma/h|$, has the
following structure. The ground-state condition of the exchange term
of the Hamiltonian imposes the restriction of having either one or two
down spins (`dimers') per triangle. Starting from a configuration
which is also a ground state of the longitudinal field part of the
Hamiltonian, we are thus allowed to add dimers violating the hardcore
condition, as long as three never meet in one site. To connect
different ground states requires relocating three dimers (denoted by
crosses in Fig.~\ref{fig:kagphase}), since the shortest closed loop of
bonds of the hexagonal lattice has length six. The lowest-order
off-diagonal matrix elements thus arise to $O((\Gamma/h)^6)$,
and are precisely those described by the quantum dimer resonance term
in Eq.~\ref{eq:triqdm}. This being the unique shortest closed loop,
all other terms up to and including sixth order are diagonal. These
terms correspond to putting down and then removing up to three
dimers. Due to the local structure of the dimer states, the number of
such operations, and the concomitant energy denominators, are found to
be the same for all ground states. Hence, the diagonal energy shift is
uniform and does not generate a lifting of the classical degeneracy.

The ordering pattern is therefore determined by the hexagonal QDM
(Eq.~\ref{eq:triqdm}) at $v=0$\ and $t\propto\Gamma(\Gamma/h)^5$. In
dimer language, it is the one depicted in Fig.~\ref{fig:triooo}.

Next, consider the kagome IAFM in a transverse field $\Gamma$ with
$h=0$.  Following our previously described strategy, we look for a
symmetry-breaking pattern within a Landau-Ginzburg-Wilson analysis;
However, this mean-field treatment predicts an infinite number of zero
modes, corresponding to the simultaneous softening of an entire branch
of excitations.\cite{reimers} For Ising spins, high-temperature
series expansion studies of the kagome IAFM indicate that thermal
fluctuations fail to select a wavevector to any
order.\cite{Harris92,Huse92}

The variational maximally flippable configurations turn out to be the
maximally polarised dimer configuration defined above for
$h>\Gamma=0$, of which there is an exponentially large number. The
resulting hierarchical states are those in which each triangle has a
spin pointing up, one pointing down and one pointing along the
transverse field. Evidently, these states map onto the three state
Potts model on the kagome lattice, which is also known to have a
nonzero entropy ${\cal S}$.

All these arguments portray the kagome TFIM as a system extremely
reluctant to order. We have checked this explicitly by quantum Monte
Carlo simulations and found that the TFIM mirrors the classical model,
with correlations somewhat enhanced compared to the classical ones,
but still rapidly exponentially decaying (see Fig.~3 of
Ref.~\onlinecite{mcs2000}). Here, we supplement these data by
displaying, in Fig.~\ref{fig:kagdetails}, that the simulated
correlation functions have settled down with respect to both quantum
temperature and discretisation error.  Note that in both cases the
correlations are extremely small in magnitude below the first few
neighbours, in marked contrast to the situation in models known to
order; for example, in the triangular IAFM the saturated correlation
function remains above $0.5$ in these units at the largest
distances.

In Fig.~\ref{fig:kaginfield}, we also show that, for a small
longitudinal field applied in addition to the transverse one, the
rapid decay of the correlation functions remains unaffected, the main
effect being the appearance of a net moment visible in the
correlations at large distances. The size of this moment is close to
linear in $h$, as one expects for a quantum paramagnet. At short
distances, we observe the emergence of weak correlations reminiscent
of the dimer crystal described above.  We therefore conjecture that
with further field-tilting results in a continuous quantum phase
transition to the ordered dimer phase. We have not been able to
confirm this numerically as our simulations fail to equilibrate before
the critical value of $h$\ is reached.

In Fig.~\ref{fig:kagphase} we display the simplest phase diagram for
the kagome IAFM in longitudinal and transverse fields consistent with
the discussion here, noting that details associated with the
tilted-field phase line remains a topic for future study.  We close
this section by restating its main result, namely the fact that the
kagome TFIM is a quantum disordered magnet.

\section{The sawtooth chain}
\label{sect:sawtooth}

An extreme and amusing example of a quantum spin liquid is provided by
the sawtooth chain (Fig.~\ref{fig:diodeconnect}). It is the ultimate
cooperative paramagnet: the locations of the frustrated bonds on each
triangle are entirely independent, providing a ground-state entropy
${\cal S}=(\ln3)/2$\ per spin. This is a result of the absence of
closed loops of triangles, and the fact that, as in the case of the
kagome lattice, they are arranged to share corners.

This chain has been studied in detail by Priour \etal\cite{donmar}
using a high-order series expansion in $J/\Gamma$, which compared
favourably with exact diagonalisations. They found no phase transition
at any value of $J/\Gamma$, implying that the chain is in a
quantum paramagnetic state.

It has been suggested that the ground-state topology plays an
important role in determining the ordering properties of a
magnet.\cite{suto} It therefore may be of interest to note that this
chain has a completely connected ground-state manifold. The proof
proceeds by explicit construction of a path from any ground state
configuration to a reference ground state configuration, pictured in
Fig.~\ref{fig:diodeconnect}. First, one picks any down spin on a
bottom row (if there are none, any one can be flipped). The remaining
two spins of the triangle can be made to agree with the reference
state by one of the three operations depicted below the reference
configuration; this is then repeated for the neighbouring triangle.
In the case of the operation pictured on the left, nothing has to be
done.  A question mark means that the spin can have either orientation
and can thus be made to agree with the reference orientation
immediately. Once this has been done, the spins enclosed by the
ellipse can be oriented appropriately.

\section{The fully frustrated hexagonal lattice}
\label{sect:ffhexagon}

\subsection{The classical model}

The fully frustrated hexagonal Ising magnet is obtained from the
ferromagnetic Ising model on that lattice by changing the sign of one
interaction in each hexagon, as depicted in
Fig.~\ref{fig:hexbasics}. Since the lattice is bond-sharing, the
classical ground states can be again be represented by hardcore dimer
coverings of the dual lattice, which is the triangular lattice. Unlike
the triangular lattice, the fully frustrated hexagonal lattice does
not admit a height model of the types encountered for other lattices
we discuss.  Although it is possible to give a rule to assign heights
to the sites of the hexagonal plaquette, these rules cannot be
consistently given for all plaquettes of the full lattice since they
involve assigning opposite orientations to neighbouring plaquettes,
which is impossible as the dual triangular lattice is not bipartite.

It is found that the classical correlations of the hexagonal FFIM are
disordered rather than critical, and the ground-state entropy of the
magnet has been evaluated to give ${\cal S}=0.214$.\cite{wolffzitt}

\subsection{Action of the transverse field and flippability analysis}

Let us now consider the dynamics induced by the transverse field. Note
that the odd coordination of the lattice precludes any spin from being
flippable individually since it cannot have an equal number of
frustrated and satisfied bonds.  The shortest allowed dimer move
consists of moving two occupied dimers and implies flipping a pair of
neighbouring spins. Not absolutely all local moves within the
ground-state manifold, however, can be generated with these pair spin
flips. In fact, one particular (`staggered') configuration exists
which does not allow any two spin flips but which instead permits the
four dimer move as depicted in Fig.~\ref{fig:hexbasics}.

For infinitesimal $\Gamma$, the ground-state degeneracy is thus lifted
to second order in $\Gamma/J$, and the relevant spin flips are the
pair flips. To identify the maximally flippable configurations, we
note that each dimer can be part of at most two flippable pairs. Since
the total number of dimers is fixed, the maximally flippable
configurations are those in which each dimer belongs to two pairs.

All maximally flippable configuration can be obtained by carrying out
any number of {\em either} operation A {\em or} B (right panel of
Fig.~\ref{fig:hexbasics}) on a particular, maximally flippable
(`columnar') configuration, in addition the symmetry operations of
global rotations and global inversion.  These operations involve
exchanging empty and occupied dimers on an infinite alternating
sequence, along a string (A) or a sawtooth (B). They generate, for a
system containing $L^2$\ spins, a number of configurations exponential
in $L$\ rather than $L^2$.  The configurations generated by operation
A all incorporate long-range order in one special direction.

\subsection{Large $\Gamma$\ analysis}

Next, we carry out the large $\Gamma$ analysis for this lattice. We
first identify the soft modes, then construct the
Landau-Ginzburg-Wilson action which we minmise in order to obtain spin
configurations and correlations functions.

%\subsubsection{Soft modes}

Since the hexagonal FFIM has four sublattices (labelled as in
Fig.~\ref{fig:hexbasics}), the Fourier transform of the interaction
matrix is of size 4x4, and is given by (omitting an overall 
factor of $J/2$: 
\bea 
%\frac{J}{2}
\left( \matrix{ 0 &
1-\expp{i q_x} & -\expp{i q_y} & 0 \cr 1-\expp{-i q_x} & 0 & 0 &
-\expp{-i q_x} \cr -\expp{-i q_y} & 0& 0& - 1-\expp{-i q_x}\cr
0&-\expp{i q_x} & - 1-\expp{i q_x}& 0 } 
\right) 
\eea 
The square of the eigenvalues, $\lambda$, are
 \bea 
%\lambda^2&=&
%\left(\frac{J}{2}\right)^2 
%\\
%&&
%\left(
3\pm
\sqrt{6+2\cos(2q_x)-2\cos(q_x+q_y)+2\cos(q_x-q_y)} 
%\right) 
\nonumber
\eea
The four extremal eigenvalues of interest are 
%$J\sqrt{6}/2$,%with factor of J/2 
$\sqrt{6}$, %without factor of J/2
which occur at
wavevectors $\pm(\pi/6,\pi/2), \pm(5\pi/6,\pi/2)$. The corresponding
eigenvectors occur in complex conjugate pairs: 
\bea 
\bv{1}&=&\bv{3}^*=
\left(\matrix{\expP{5}{12}/F\cr\expP{-}{6}/F
\cr\expP{-}{12}\cr1}\right) \exp\left(\frac{\pi i}{6}x+\frac{\pi
i}{2}y \right)\\ \bv{2}&=&\bv{4}^*=
\left(\matrix{\expP{}{12}\cr\expP{-5}{6}
\cr\expP{-5}{12}/F\cr1/F}\right) \exp\left(\frac{\pi i}{6}x+\frac{\pi
i}{2}y \right), 
\eea where $F=2\sin(5\pi/12)$.

In order to determine the terms in the LGW Hamiltonian, one determines
how these modes transform among themselves under the symmetry
operations (translations (${\rm T_x, T_y}$), reflection (${\rm Ref}$)
and rotation (${\rm Rot}$)) of the underlying lattice. Each of these
symmetry operations comes with a gauge transformation since the unit
cell of the lattice and the unit cell of the interactions are
unequal. These are shown in Fig.~\ref{fig:hexlgw}. The transformation
matrices for the amplitudes of these modes under the abovementioned
symmetry operations are thus:

\bea
{\rm T_x}&=& \left(\matrix{ {e^{{\fracc{i\pi}{6}} }} & 0 & 0 & 0 \cr 0
	& {e^ {{\fracc{5\,i\pi}{6}} }} & 0 & 0 \cr 0 & 0 & {e^
	{{-\fracc{i\pi}{6}} }} & 0 \cr 0 & 0 & 0 & {e^
	{{-\fracc{5\,i\pi}{6}} }} \cr } \right)\nonumber\\
{\rm T_y}&=& \left(\matrix{ 0 & 0 & 0 & {e^{{-\fracc{i\pi}{12}} }} \cr
	0 & 0 & {e^ {{-\fracc{5\,i\pi}{12}} }} & 0 \cr 0 & {e^
	{{\fracc{i\pi}{12}} }} & 0 & 0 \cr {e^
	{{\fracc{5\,i\pi}{12}} }} & 0 & 0 & 0 \cr } \right)\nonumber\\
{\rm Ref}&=& \frac{1}{\sqrt{2}}\left(\matrix{ 0 & i & {e^{{{i}{}}\pi/4
	}} & 0 \cr -i & 0 & 0 & {e^ {{\fracc{i\pi}{4}} }} \cr
	{e^{{-\fracc{i\pi}{4}} }} & 0 & 0 & -i \cr 0 &
	{e^{{-\fracc{i\pi}{4}} }} & i & 0 \cr }\right) \nonumber\\
{\rm Rot}&=&\frac{1}{\sqrt{2}}\left(\matrix{ 0 &
     {e^{{\fracc{i\pi}{12}} }} & {e^ {{-\fracc{i\pi}{6}} }} & 0 \cr
     {e^ {{-\fracc{7\,i\pi}{12}} }} & 0 & 0 & {e^{{\fracc{i\pi}{6}}
     }} \cr {e^{{\fracc{i\pi}{6}} }} & 0 & 0 & {e^
     {{-\fracc{i\pi}{12}} }} \cr 0 & {e^{{-\fracc{i\pi}{6}} }} & {e^
     {{\fracc{7\,i\pi}{12}} }} & 0 \cr } \right) \nonumber
\eea

To find the eventual action, one has to determine the terms at each
order which remain invariant under these transformations. One obtains:

\bea { \cal{L}}&=&
	(r+q^2)(\psi_a^2+\psi_b^2)+u_4(\psi_a^2+\psi_b^2)^2+
	\nonumber\\&&+u_6(\psi_a^2+\psi_b^2)^3+\nonumber\\&&
	+v_6(\psi_a\psi_b^5\cos(\theta_a-5\theta_b)+
	\psi_a^{5}\psi_b\cos(5\theta_a-\theta_b))\nonumber
	\eea

Here we have transformed the four mode amplitudes $\{a_{i}|i=1...4\}$\
into complex numbers $\psi_{a,b}\exp({i \theta_{a,b}})$\ with
$\psi_{a}\exp({i\theta_{a}})=a_1+i a_3$\ and
$\psi_{b}\exp({i\theta_{b}})=a_2+i a_4$.

The terms which survive are $O(4)$-symmetric up to sixth order, where
in addition a symmetry-breaking term appears. To minimise this action,
we use the $O(4)$\ symmetric terms to determine $\psi_a^2+\psi_b^2$,
the radius $R$\ of the $O(4)$-sphere, which sets the overall amplitude
of the spin pattern. The symmetry-breaking $v_6$-term then selects 48
points on this sphere. One finds that $\psi_a=\psi_+$\ and
$\psi_a=\psi_+/F$; $\theta_{a,b}=0\,\,\, (\pm \pi/6)$\ for
$v_6<\,\,(>)\,\,0$. The other 47 solutions are generated by the
operations $\psi_a\leftrightarrow\psi_b$\ and
$\{\theta_a\rightarrow\theta_a+\pi/12,\,\,
\theta_b\rightarrow\theta_b+5\pi/12\}$.  The 48 solutions for a given
sign of $v_6$\ are related by simple symmetry operations. Note the
large unit cell of the ordering pattern, which contains 48 spins. This
can in part be attributed to the non-uniform pattern of bonds. In the
gauge-invariant dimer language, the unit cell could be smaller; an
example of this is given further down.

The correlation functions obtained from the above expressions are for
soft spins so that they will certainly not be quantitatively found in
the real system at low temperatures. However, their qualitative
features, if the ordering pattern is correctly predicted, should
survive. These are peaks at the appropriate wavevectors in Fourier
space; due to the factors of $1/F$\ for the amplitudes $\psi_{a,b}$,
these peaks are not expected to have equal heights for correlation
functions for spins on the same sublattice. It turns out that the
calculated correlation functions when, averaged over all 48 minima,
are the same as those averaged over the entire $O(4)$\ sphere. 

We have looked for this ordering pattern by quantum Monte Carlo. There
are several features complicating this search. One arises from the
quantum dynamics which takes the form of double spin flips. Luckily,
one is saved here by the fact that the cluster algorithm can be
generalised to this case. The most simple-minded extension to double
spin flips would have been to generate two neighbouring clusters (rods
in the imaginary time direction) separately -- by design, the cluster
algorithm would cancel the problematic Boltzmann factor
$\exp(-K^\tau/2)$\ in each of the two clusters separately. However,
clusters of unequal height, which become overwhelmingly probably for
large $K^\tau$, cannot be flipped because permissible moves must be
double spin flips in any and all planes. The way around this is to
construct a cluster as a pair of rods. This works because the cluster
can either terminate when encountering zero, one or two domain walls
in the time direction. The case of zero and two can be taken care of
by working with an effectively doubled coupling in the time direction;
the case of one domain wall takes care of itself as there is no
Boltzmann factor to be canceled in the first place. This prescription
can therefore not be generalised to other multiple-spin flips in a
straightforward manner.

Another problem is the very large unit cell of the ordering pattern,
which is 48 spins but only fits into periodic boundary conditions for
system sizes multiples of 192 spins. We have thus not attempted to do
a complete finite-size scaling study and have contented ourselves with
displaying the presence of peaks at the wavevectors predicted
above. 

In Fig.~\ref{fig:hexmc} we have displayed the correlation function for
spins on sublattice 4 along the $x$\ and $y$\ directions (also see
Fig.~\ref{fig:hexbasics}). A peak at the expected location is clearly
visible. This suggests that there is at least significant short-range
order of the predicted nature present in this regime. We caution,
however, that we have not been able to do a complete numerical
analysis of this problem that would unambiguously establish this
ordering pattern.

To rationalise the above results, we finally present a dimer pattern
which corresponds to a spin pattern of the correct size unit cell
(Fig.~\ref{fig:mcgov}).\cite{husedis} We have not found a way of
relating the spin-spin correlations precisely to those of the
solutions of the LGW theory found above, but one can see rather nicely
how the periodicity of twelve and four in the $x$\ and $y$\ directions
comes about from a smaller size unit cell in the gauge invariant
description. The state pictured is favourable because each spin
fluctuates in a pair gaining energy from the transverse field, and
because there are many equivalent such configurations, as explained in
the caption.

\section{Summary}
In this paper we have described our analysis of a number of frustrated
Ising systems with a quantum dynamics introduced by a magnetic field
applied transverse to the Ising axis. We have argued that these models
are of theoretical interest as the simplest settings in which quantum
dynamics interacts with classical frustration. They can represent
effective
theories of systems, such as short ranged RVB magnets, where the low
energy dynamics contains a frustrated Ising degree of freedom and are
likely of experimental interest when suitable materials are probed by
the application of a transverse magnetic field.       

We have been able to make considerable progress in elucidating the
phase structure of these systems which sheds much light on the
interplay between the structure of the macroscopic degeneracy
and the quantum dynamics.
Our results include instances of ``order by disorder'' and of
``disorder by disorder'' (i.e. Ising spin liquids). In arriving at      
these we have used two systematic approaches: a variational approach
that builds on the local entropy of different classical configurations
that is the Ising analog of semiclassical analyses of magnets with   
continuous symmetry, and an LGW analysis that attempts to guess at
the large order structure of an expansion about the large transverse field
paramagnet. We have also made use of quantum Monte Carlo
simulations as well as of mappings to height and dimer models
with the latter allowing a connection to RVB physics.

There appear to be several directions that can be pursued further.
Along the lines of the questions addressed in this paper, the analysis of 
the kagome system in tilted fields and a definitive analysis
of the hexagonal lattice problem are needed. Further, he structure and 
energetics of the low energy excitations needs to be worked out to have an 
understanding of the dynamics at low temperatures. Beyond this it would
be interesting to include the dynamics of the transverse exchange and
see what happens both in the approach to the Heisenberg limit and when
a transverse field is also present. We expect such work to be both
fruitful and instructive.

\section*{Acknowledgements}
We are very grateful to P. Chandra for collaboration on many aspects
of this work. We would also like to thank M. Aizenman, J. Chalker,
C. Henley, D. Huse, S. Kivelson, E. Lieb and S. Sachdev for useful
discussions. This work was supported in part by grants from the
Deutsche Forschungsgemeinschaft, the NSF (grant No. DMR-9978074), the
A. P. Sloan Foundation and the David and Lucille Packard Foundation.

\newpage
\mbox{}
\newpage

\begin{figure}
\epsfxsize=3.4in
\centerline{\epsffile{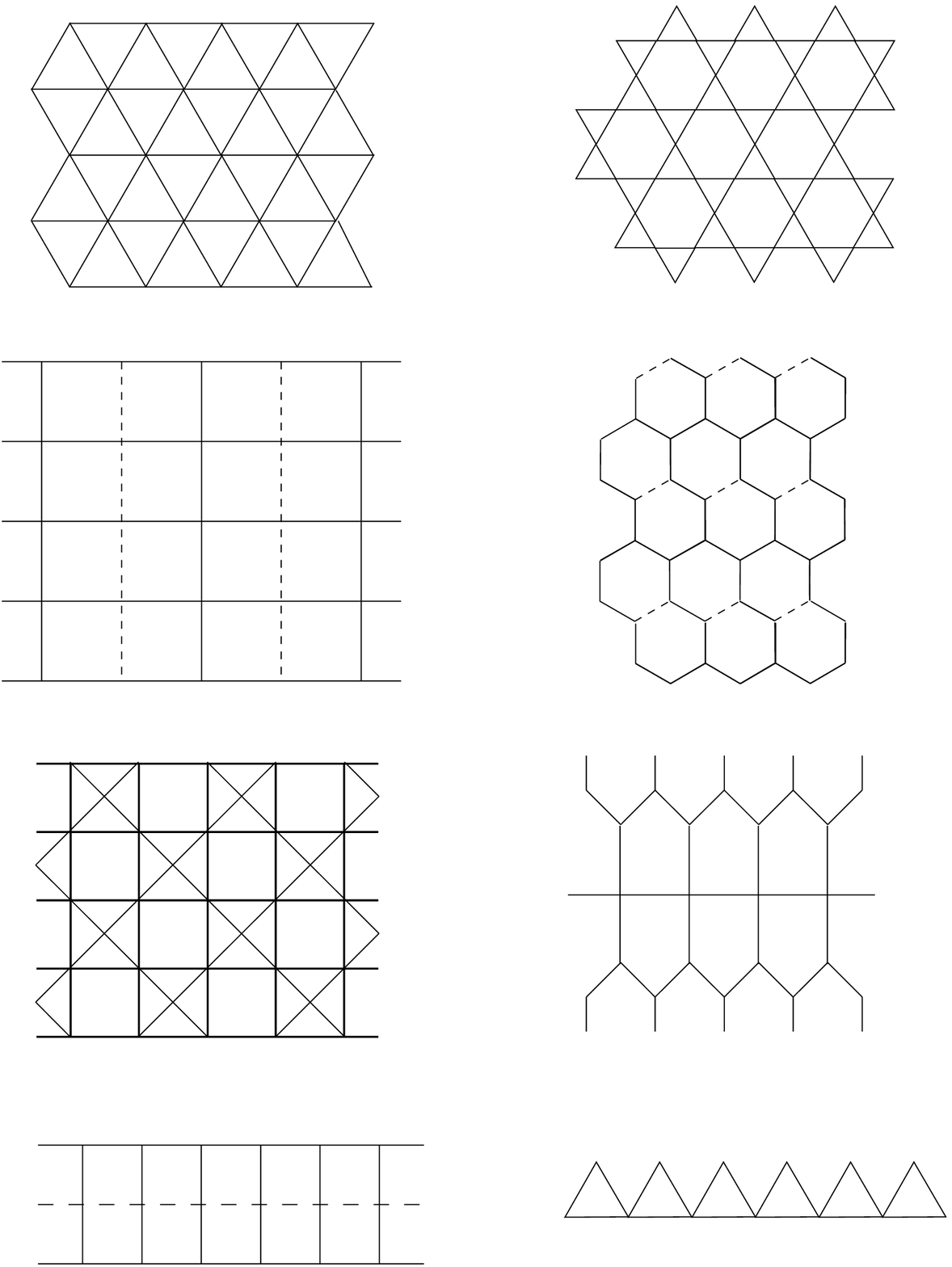}}
\caption{ The lattices on which fully frustrated transverse field
Ising models are discussed in this paper. Clockwise from noon: kagome,
hexagonal, pentagonal, sawtooth chain, three-leg ladder, square
lattice with crossings (``two-dimensional pyrochlore''), square and
triangular. For normally unfrustrated lattices, frustration is
introduced by choosing an odd number of bonds in each plaquette to be
antiferromagnetic (see Fig.~\ref{fig:frusplaq}).}
\label{fig:alllattices}
\end{figure}

\begin{figure}
\epsfxsize=3.2in
\centerline{\epsffile{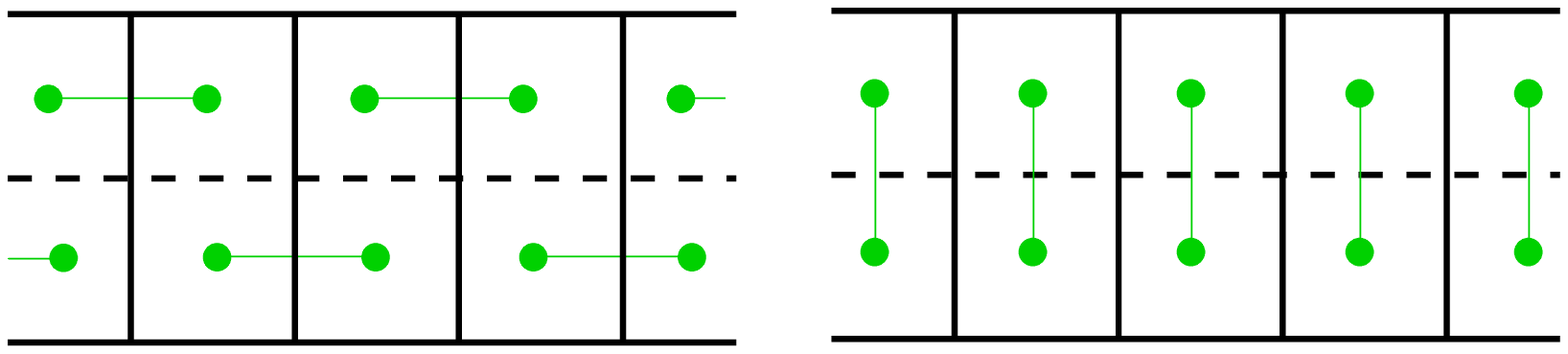}}
\caption{ The fully frustrated three-leg ladder. Solid lines
correspond to ferromagnetic bonds, dashed lines to antiferromagnetic
ones. Dimer representations of the columnar (right) 
and one of the two
staggered (left) configurations.}
\label{fig:fftll}
\end{figure}

\begin{figure}
\epsfxsize=3.2in
\centerline{\epsffile{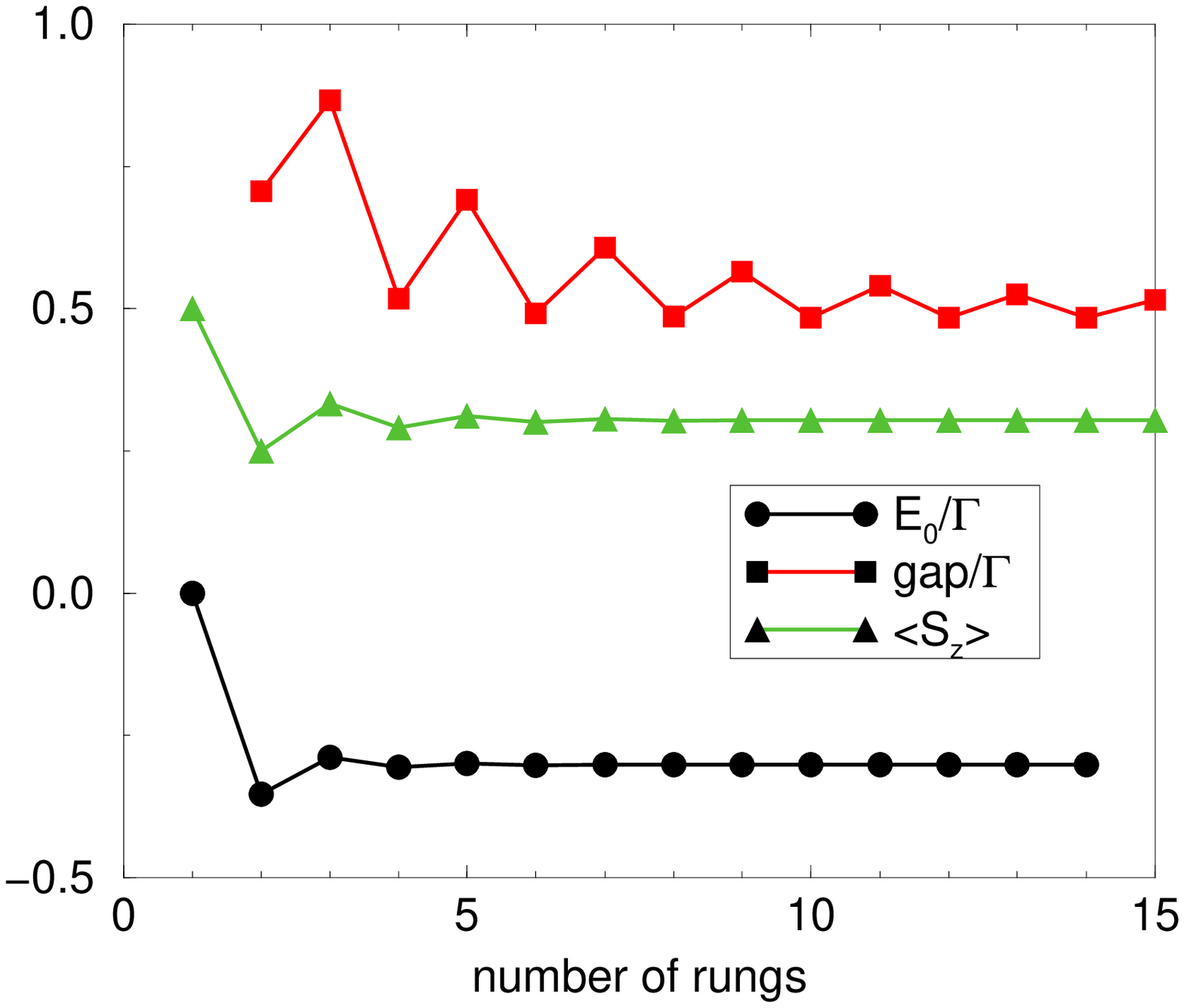}}
\caption{ Ground-state energy (proportional to $\left<S_x\right>$),
energy gap and $\left<S_z\right>$ for the three-leg lader in the limit
$\Gamma\rightarrow 0^+$\ from exact diagonalisation.  }
\label{fig:tllplot}
\end{figure}

\begin{figure}
\epsfxsize=3.2in
\centerline{\epsffile{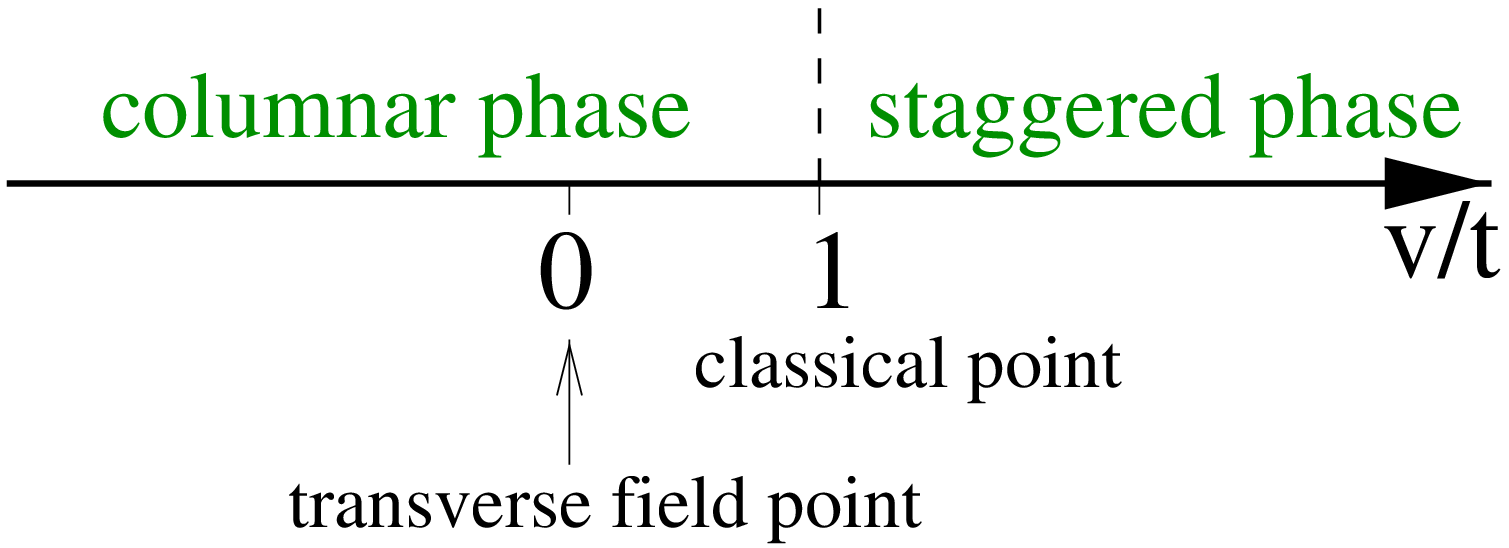}}
\caption{ The phase diagram of the Rokhsar-Kivelson quantum dimer
model for the fully frustrated Ising magnet on the three-leg
ladder. The diagonal correlations at $v=t$\ are those of the classical
dimer model. The infinitesimal transverse field problem, $\Gamma=0^+$,
maps to the point $v=0$ and thus corresponds to a finite jump of $v$.}
\label{fig:rkphase}
\end{figure}

\begin{figure}
\epsfxsize=2.0in
\centerline{\epsffile{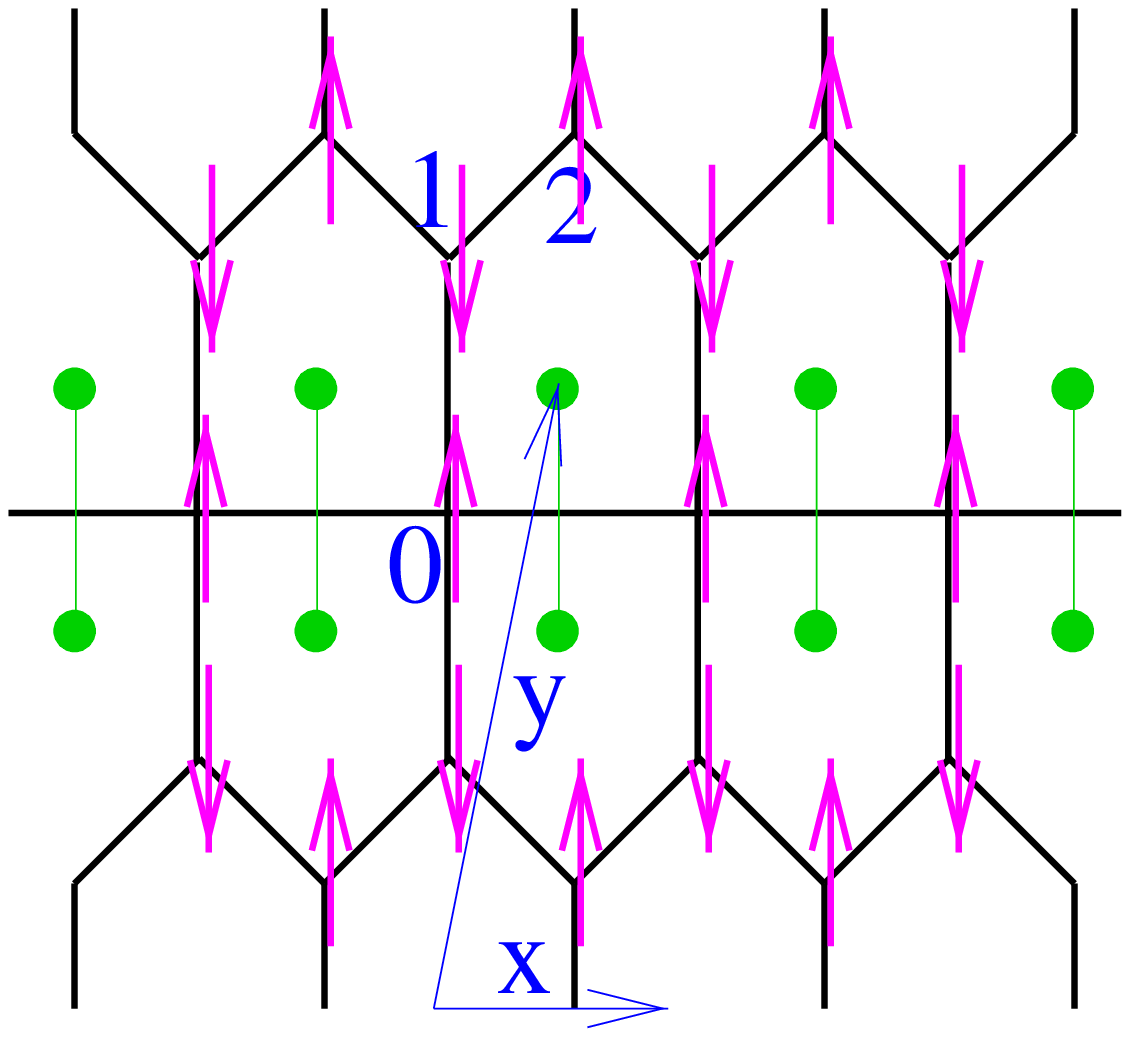}}
\caption{The pentagonal lattice with its maximally flippable
configuration. The three spins in the unit cell are labelled by
(0,1,2), and the $x$\ and $y$-lattice translation vectors are given by
the arrows.  The leading-order perturbation theory reduces to studying
the central horizontal ladder, which is equivalent to the three-leg
ladder of Fig.~\ref{fig:fftll}}
\label{fig:pent}
\end{figure}

\begin{figure}
\epsfxsize=3.2in
\centerline{\epsffile{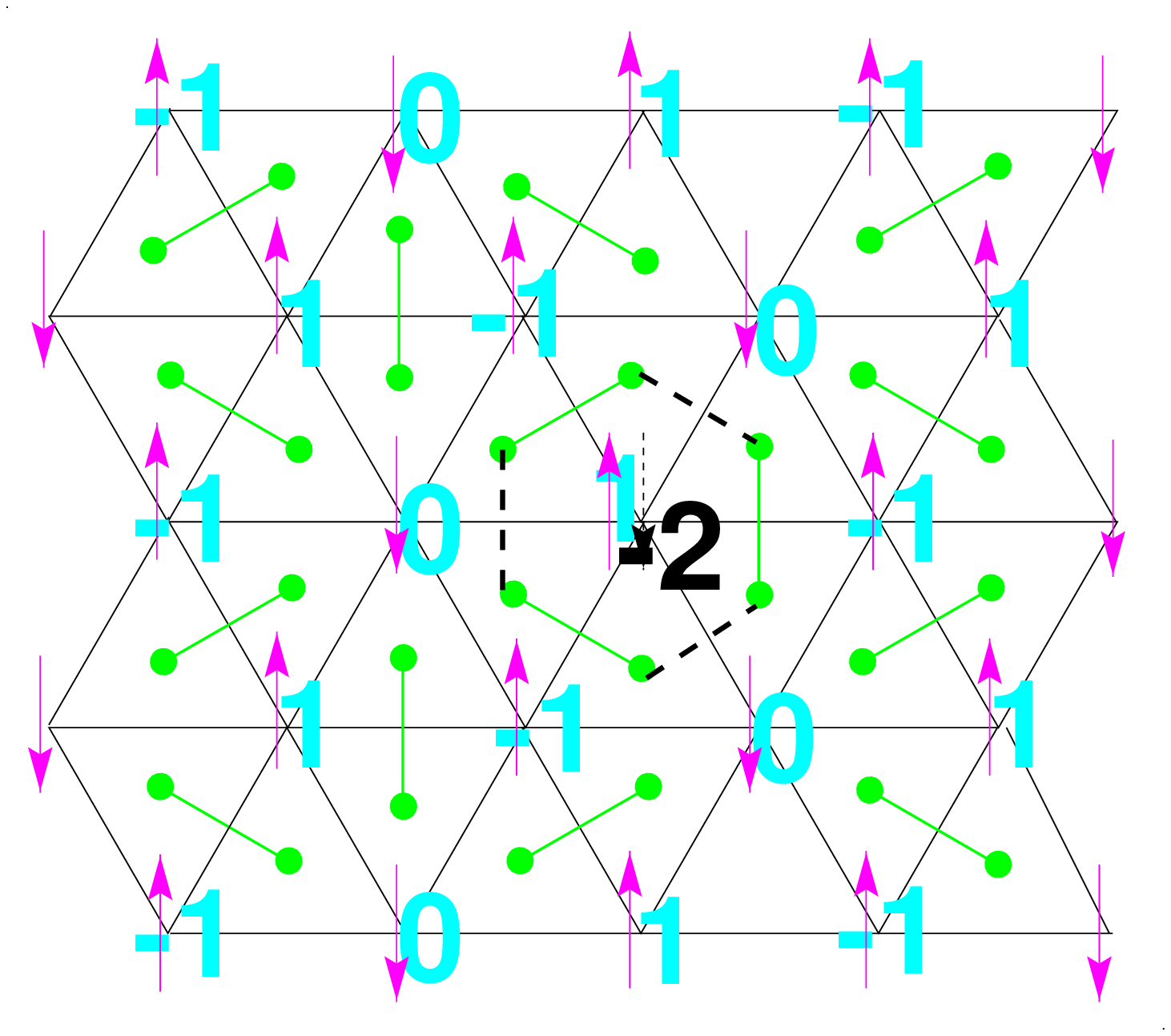}}
\caption{ The maximally flippable spin state (arrows) and its
corresponding height and dimer configurations on the triangular
lattice. The result of a single spin flip for the dimer and height
configurations is also shown.  See the text for details. }
\label{fig:triooo}
\end{figure}

\begin{figure}
\epsfxsize=2.2in
\centerline{\epsffile{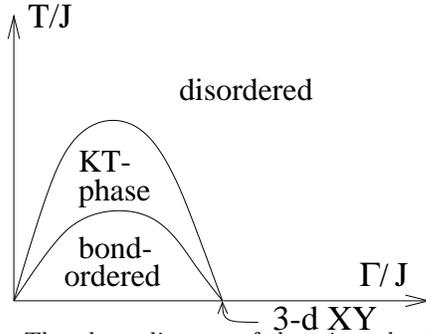}}
\caption{ The phase diagram of the triangular TFIM. The pattern of the 
ordered phase is depicted in Fig.~\ref{fig:triooo}}
\label{fig:jose}
\end{figure}

\begin{figure}
\epsfxsize=3.5in
\centerline{\epsffile{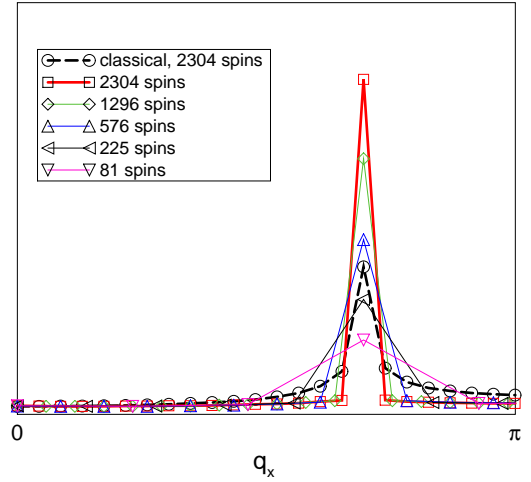}}
\caption{ The absolute value of the Fourier transform of the spin-spin
correlation function along the $x$-direction, $\langle
S(x)S(0)\rangle$, (arbitrary units, at low temperature), as a function
of $q_x$\ ranging from 0 to $\pi$. Note the growing and narrowing
Bragg peak.}
\label{fig:tribragg}
\end{figure}

\begin{figure}
\epsfxsize=3.5in
\centerline{\epsffile{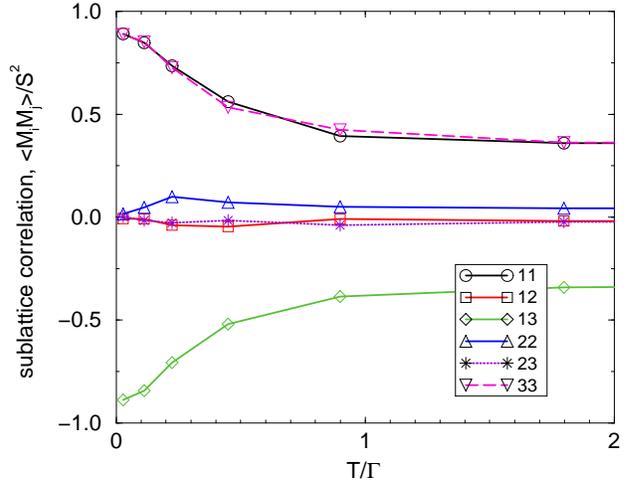}}
\caption{ The correlation matrix,
$\langle M_iM_j\rangle=\langle M_jM_i\rangle$ of sublattice
magnetisations, $M_i$, for a three sublattice state, as a function of
the quantum temperature for a system with 729 spins. The sublattices
are labelled by the size of their magnetisation so that
$M_1>M_2>M_3$. $\left<M_2^2\right>\simeq0$ implies that one sublattice
has zero root-mean square magnetisation.
$\left<M_1^2\right>\simeq\left<M_3^2\right>\simeq-\left<M_1M_3\right>$\
implies the other two sublattices have equal and opposite
magnetisation.}
\label{fig:triagnew}
\end{figure}

\begin{figure}
\epsfxsize=3in
\centerline{\epsffile{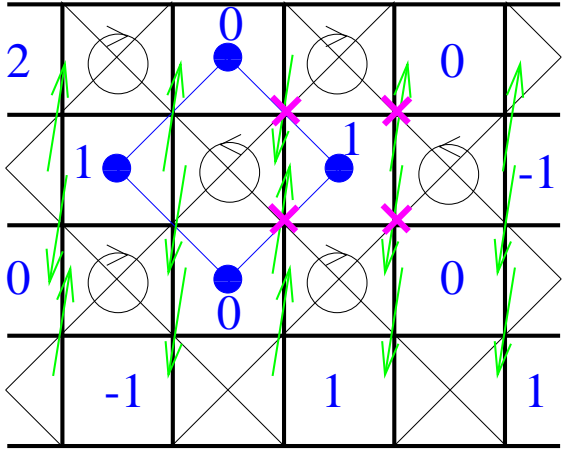}}
\caption{ The square lattice with crossings, or two-dimensional
pyrochlore.  A spin and the resulting height configuration are
shown. Other features are explained in the text.}
\label{fig:sqcrossheight}
\end{figure}

\begin{figure}
\epsfxsize=2.2in
\centerline{\epsffile{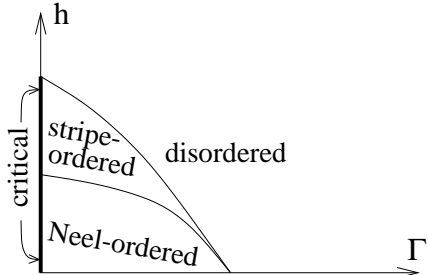}}
\caption{ The phase diagram in the $h-\Gamma$\ plane for the square
lattice with crossings. The precise extent of either phase away from
$\Gamma/h\ll 1$\ is unknown.}
\label{fig:sqcrossphase}
\end{figure}

\begin{figure}
\epsfxsize=3in
\centerline{\epsffile{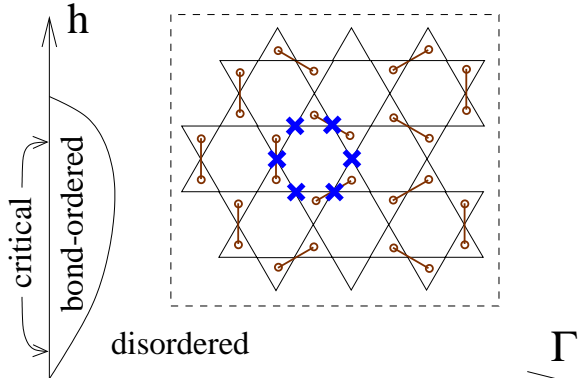}}
\caption{ Phase diagram for the kagome Ising antiferromagnet in a
field. Inset: Mapping of the kagome IAFM in a longitudinal field onto
the hexagonal lattice dimer model. 
The down spins are marked by dimers. The
non-trivial move to sixth order in perturbation theory corresponds to
flipping all the spins marked by crosses.  }
\label{fig:kagphase}
\end{figure}

\begin{figure}
\epsfxsize=3.2in
\centerline{\epsffile{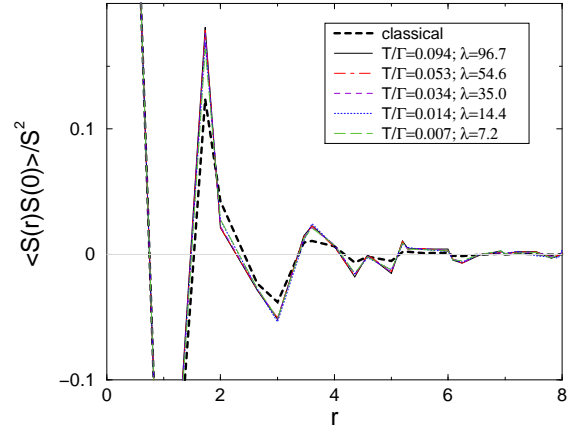}}
\caption{ The angularly averaged spin-spin correlation function for
different quantum temperatures and discretisations,
$\lambda=\exp(K^\tau/2)$. Note the rapid decay and the enlarged
scale. The quantum curves lie almost on top of one another.}
\label{fig:kagdetails}
\end{figure}

\begin{figure}
\epsfxsize=3.2in
\centerline{\epsffile{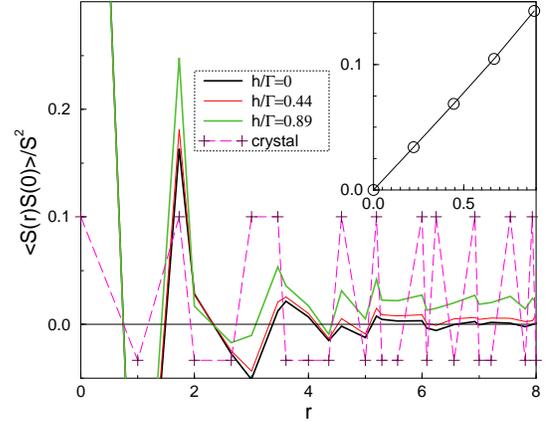}}
\caption{ The angularly averaged spin-spin correlation function for
the kagome TFIM in a longitudinal field. The correlations
corresponding the dimer crystal (Fig.~\ref{fig:kagphase}), scaled down
by a factor of 10, are also shown. Note that the high-field curve
starts moving towards the crystal correlations at small distances.
Inset: The magnetisation per spin (ordinate) is linear in field
($h/\Gamma$, abscissa).}
\label{fig:kaginfield}
\end{figure}

\begin{figure}
\epsfxsize=2.2in
\centerline{\epsffile{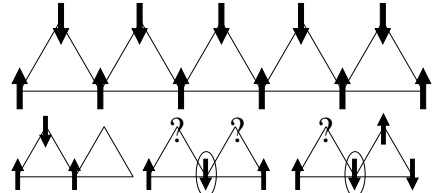}}
\caption{ The sawtooth chain with the reference ground state and the
three operations needed 
for proving ground-state connectedness (see text).  }
\label{fig:diodeconnect}
\end{figure}

\begin{figure}
\epsfxsize=3.0in
\centerline{\epsffile{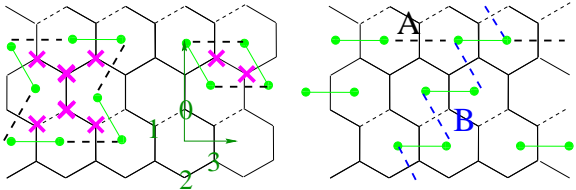}}
\caption{ The fully frustrated hexagonal lattice. The basis spins of
the lattice are numbered 0-3 and the rectangular lattice translation
vector in the $x$\ $(y)$\ direction is given by the horizontal
(vertical) arrow. The pair spin flip and a non-trivial multiple spin
flip are indicated with their corresponding dimer moves. Maximally
flippable states are generated from the columnar seed state by
interchanging occupied and empty dimers along horizontal lines (A) or
in a sawtooth pattern (B). }
\label{fig:hexbasics}
\end{figure}

\begin{figure}
\epsfxsize=3.0in
\centerline{\epsffile{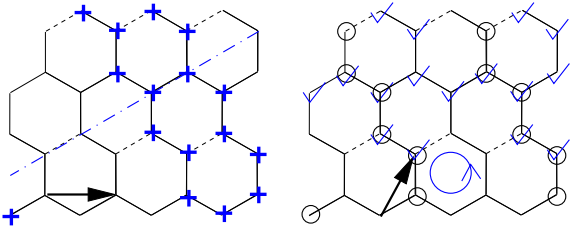}}
\caption{The symmetry transformations and concomitant gauge
transformations for the Landau-Ginzburg-Wilson action. Translation in
the $x$-direction (arrow in left panel) require no gauge
tranformation. Reflections about the dot-dashed line go along with
inversion of the spins marked by crosses; translations by the other
hexagonal lattice vector (arrow in the right panel) require flipping
the circled spins, and rotations by $\pi/3$ (circle with arrow)
involve flipping the spins denoted by ticks.}
\label{fig:hexlgw}
\end{figure}

\begin{figure}
\epsfxsize=3.5in
\centerline{\epsffile{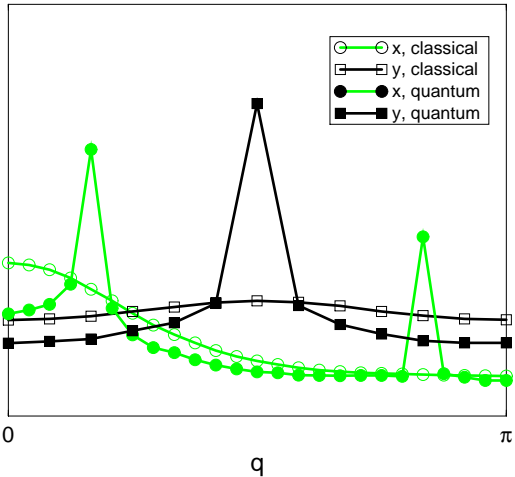}}
\caption{ The Fourier transform of the absolute value of the
autocorrelation function of a spin on sublattice 4 in both $x$\ and
$y$ directions for a system of 4608 spins.  Results for the classical
and the quantum ($\Gamma=0^+$) cases are shown, with a quantum
temperature $T_Q/\Gamma=0.028$\ nominally, but the discretisation
error is rather large: $\lambda=4.5$\ only.}
\label{fig:hexmc}
\end{figure}

\begin{figure}
\epsfxsize=3.0in
\centerline{\epsffile{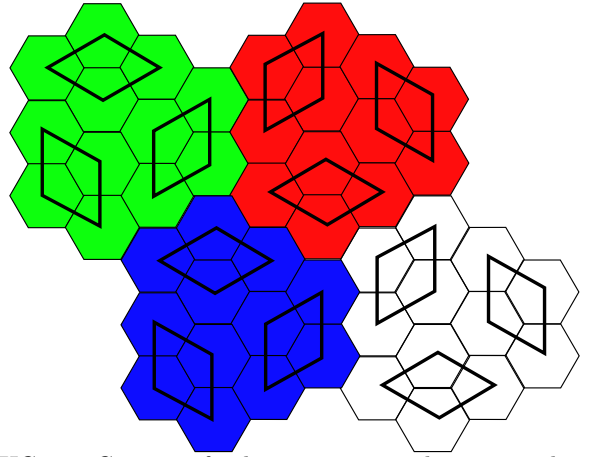}}
\caption{Cartoon of a dimer pattern with a maximal number of
independently flippable pairs of spins. To obtain the corresponding
spin pattern, one has to choose a gauge, i.e.\ one has to make a
choice of which bonds are antiferromagnetic. Fig.~\ref{fig:hexbasics}
shows one such choice. For this choice, the spin pattern has a 48 site
unit cell (two plaquettes on the left), whereas the gauge invariant
dimer description only has 24 sites in a unit cell (any single
plaquette).  A rhombus stands for a pair of dimers (frustrated bonds)
resonating between two configurations, so that the two spins enclosed
by it can be thought of as flipping together to take advantage of the
transverse field.  This pattern cannot be derived from a uniform
maximally flippable state but has extensive `configurational' entropy,
as there are two distinct ways (shown in the left and right pairs of
plaquettes, respectively) of pairing up the spins in each plaquette.
}
\label{fig:mcgov}
\end{figure}


\begin{references}

\newpage

\bibitem{wannier} G. H. Wannier,
\tit{\pr}{79}{357}{1950}{Antiferromagnetism. The triangular Ising
net}.

\bibitem{houtappel} R. M. F. Houtappel,
\tit{Physica\ }{16}{425}{1950}{Order-disorder in hexagonal lattices}.


\bibitem{fn-caveat1} In truth, the issue is somewhat delicate in this
particular case, with the zero temperature limit of the spin
correlations being critical.\cite{stephensontri} A better example is the
kagome lattice Ising antiferromagnet (Fig.~\ref{fig:alllattices}); we
will return to these subtleties below.

\bibitem{stephensontri}
J. Stephenson, \jour{\jmp}{11}{413}{1970}; 
\jour{\jmp}{11}{420}{1970}.


\bibitem{villain}
J. Villain, \tit{\zpb}{33}{31}{1979}{Insulating Spin Glasses}.

\bibitem{hfmrev}
For a review, see: R. Moessner, cond-mat/0010301.

\bibitem{fn-frus} We should note that there is considerable laxity in
the literature regarding the meaning of the term ``frustrated magnet''
even for classical systems.  It would be best to restrict it to
systems that have an extensive entropy or ground state dimensionality
at $T=0$\ but it is not unusual to use it to describe all systems with
bond interactions that are not simultaneously satisfiable. An instance
of the latter, but not of the former is the triangular lattice
Heisenberg model. We should also note that even with the more
restricted usage it is not synonymous with absence of ordering -- a
constrained four states Potts model on the square lattice
(J. K. Burton and C. L. Henley, cond-mat/9708171) appears to be an
example of an ordered system with extensive entropy at $T=0$, as is a
three-leg ladder that we discuss below.

\bibitem{Huse92}
D.A. Huse and A.D. Rutenberg, \jour{\prb}{45}{7536}{1992}.

\bibitem{Villain80} 
J. Villain, R. Bidaux, J. P. Carton and
R. J. Conte, \tit{J. Phys. -- Paris\ } {41}{1263}{1980}{Order as an
effect of disorder}.

\bibitem{shenderquantum}
E. F. Shender, 
\tit{Sov. Phys. JETP\ }{56}{178}{1982}{Anti-ferromagnetic
garnets with fluctuationally interacting sub-lattices}.

\bibitem{collinsrev} For a review, see: M. F. Collins and
O. A. Petrenko, \tit{Can.\ J.\ Phys.\ }{75}{605}{1997}{Triangular
antiferromagnets}.

\bibitem{kagdiag} see, e.g., J. T. Chalker and J. F. G. Eastmond,
\tit{\prb}{46}{14201}{1992}{Ground-state disorder in the spin-1/2
kagome heisenberg-antiferromagnet}; 
C. Zeng and V. Elser,
\tit{\prb}{51}{8318}{1995}{QUANTUM DIMER CALCULATIONS ON THE SPIN-1/2
KAGOME HEISENBERG-ANTIFERROMAGNET}; 
P. Lecheminant, B. Bernu,
C. Lhuillier, L. Pierre and P. Sindzingre,
\tit{\prb}{56}{2521}{1997}{Order versus disorder in the quantum
Heisenberg antiferromagnet on the kagome lattice using exact spectra
analysis}; F. Mila, \tit{\prl}{81}{2356}{1998}{Low-energy sector of
the S=1/2 Kagome antiferromagnet} and references therein.

\bibitem{Fazekas74}
P. W. Anderson, Mat.\ Res.\ Bull.\ {\bf 8}, 153 (1973);
P. Fazekas and P. W. Anderson, {Phil.\ Mag.\ }{\bf 30}, 23 (1974).

\bibitem{fn-extensions} In the future we hope to explore other
instances of single spin quantum dynamics involving either larger
values of spin or the extension of classical continuous spin models to
quantum rotor models. The quantum dynamics of mobile holes in a
frustrated Ising background is discussed in Ref.~\onlinecite{MSnag}.

\bibitem{MSnag}
R. Moessner and
S. L. Sondhi, { to appear in Phys.\ Rev.\ B.}

\bibitem{mcs2000} 
R. Moessner, S. L. Sondhi and P. Chandra,
\tit{\prl}{84}{4457}{2000}{Two-dimensional periodic frustrated Ising
models in a transverse field}.

\bibitem{Chakrabarti96}
For some work on one-dimensional models, see
B.K. Chakrabati, A. Dutta and P. Sen, {\sl Quantum
Ising Phases and Transitions in Tranverse Ising Models},
(Springer-Verlag, Berlin, 1996).

\bibitem{Sachdev99}
S. Sachdev, {\sl Quantum Phase Transitions}
(Cambridge University, Cambridge, 2000).

\bibitem{rsrg}
D. S. Fisher,
\jour{\prl}{69}{534}{1992}.

\bibitem{Aeppli98}
G. Aeppli and T.F. Rosenbaum in {\sl Dynamical Properties of
Unconventional
Magnetic Systems}, A.R. Skjeltorp and D. Sherrington eds., 
(Kluwer Academic Publishers, Amsterdam, 1998).

\bibitem{pwa87}
P. W. Anderson, \tit{\sci}{235}{1196}{1987}{THE RESONATING VALENCE
BOND STATE IN LA2CUO4 AND SUPERCONDUCTIVITY}.

\bibitem{Rokhsar88}
D. S. Rokhsar and S. A. Kivelson,
\tit{\prl}{61}{2376}{1998}{SUPERCONDUCTIVITY AND THE QUANTUM HARD-CORE
DIMER GAS}.

\bibitem{readsach} 
N. Read and S. Sachdev, 
\jour{\npb}{316}{609}{1989}; \jour{\prl}{66}{1773}{1991}.

\bibitem{wen} X.G. Wen, 
\jour{\prb}{44}{2664}{1991}.

\bibitem{mudry94} 
C. Mudry and E. Fradkin, \tit{\prb}{49}{5200}{1994}{
Separation of Spin and Charge Quantum Numbers in Strongly Correlated
Systems};
\tit{\prb}{50}{11409}{1994}{The Problem of Separation of
Spin and Charge in One-Dimensional Quantum Antiferromagnets}.


\bibitem{SV} S. Sachdev and M. Vojta,
\tit{\jpsj}{69}{1}{2000}{Translational symmetry breaking in
two-dimensional antiferromagnets and superconductors} 

\bibitem{SF}
T. Senthil and M. P. A. Fisher, 
\tit{\prb}{62}{7850}{2000}{Z(2) gauge theory of electron 
fractionalization in strongly correlated systems}.

\bibitem{MStrirvb}
R. Moessner and S. L. Sondhi,  {cond-mat/0007378}.

\bibitem{Suzuki71} M. Suzuki, { Prog. Theor. Phys.} {\bf 46}, 1337,
(1971); {\sl ibid} {\bf 56}, 2454 (1976).

\bibitem{Trotter59} 
H.F. Trotter, { Proc. Am. Math. Soc.} {\bf 10},
545 (1959).

\bibitem{ferswen} R. H. Swendsen, J. S. Wang and A. M. Ferrenberg,
\tit{Top.\ Appl.\ Phys.\ }{71}{75}{1992}{NEW MONTE-CARLO METHODS FOR
IMPROVED EFFICIENCY OF COMPUTER- SIMULATIONS IN STATISTICAL-MECHANICS
}

\bibitem{riegercont}
H. Rieger and N. Kawashima, \tit{\epjb}{9}{233}{1999}{Application 
of a continous time cluster algorithm to the
          two-dimensional random quantum Ising ferromagnet}

\bibitem{fn-dimtop} We note that the existence of a dimer mapping has
interesting topological consequences,\cite{Rokhsar88} which however
will not be discussed in this work.


\bibitem{husedis}
We thank David Huse for a useful  discussion of this point.

\bibitem{coppersmith} This is related to the fact that in a
layered, ferromagnetically stacked frustrated magnets, fluctuations
will determine which state is selected as ground state in the limit
$T\rightarrow 0$. The importance of fluctuations for such stacked
magnets was pointed out by 
S. N. Coppersmith, \tit{\prb}{32}{1584}{1985}{LOW-TEMPERATURE 
PHASE OF A STACKED TRIANGULAR ISING ANTIFERROMAGNET}. In our
formulation, fluctuations are naturally included, and the mapping to
the stacked magnets provides a formal connection
between these results.

\bibitem{fn-hopping} More precisely, the graph on which hopping is
occurs can be defined as a high dimensional but finite lattice.  For a
system of $N$\ spins, it is a subgraph of an $N$\ dimensional
hypercube.

\bibitem{pyroshlo} 
R. Moessner and J. T. Chalker,
\tit{\prl}{80}{2929}{1998}{Properties of a classical spin liquid: The
Heisenberg pyrochlore antiferromagnet};
\tit{\prb}{58}{12049}{1998}{Low-temperature properties of classical
geometrically frustrated antiferromagnets}.

\bibitem{Blankschtein84}
D. Blankschtein, M. Ma, A.N. Berker, G.S. Grest and C.M. Soukoulis,
\tit{\prb}{29}{5250}{1984}{ORDERINGS OF A STACKED FRUSTRATED 
TRIANGULAR SYSTEM IN 3 DIMENSIONS}.

\bibitem{blankvillain}
D. Blankschtein, M. Ma and A.N. Berker, 
\tit{\prb}{30}{1362}{1984}{FULLY AND PARTIALLY FRUSTRATED 
SIMPLE-CUBIC ISING-MODELS - LANDAU-GINZBURG-WILSON THEORY}.

\bibitem{donmar}
D.J. Priour Jr, M. P. Gelfand and S. L. Sondhi, cond-mat/0005185.

\bibitem{fn-nosym} Note that the columnar dimer state is disordered in
dimer language as it breaks no symmetries of the lattice. Had the
hierarchical state been selected, there would have been breaking of 
translational symmetry.

\bibitem{pent} M. H. Waldor, W. F. Wolff and J. Zittartz,
\jour{\pla}{106}{261}{1984}; \jour{\zpb}{59}{43}{1985}.

\bibitem{Blote82}
H.W.J. Bl\"{o}te and H.J. Hilhorst, {\sl J. Phys. A}{\bf 15}, L631
(1982);
B. Nienhuis, H.J. Hilhorst and H.W. Bl\"{o}te, {\sl ibid} {\bf 17}, 3559
(1984).

\bibitem{fn-Orland} We note that this specific problem has been
treated by Orland (P. Orland, \tit{\prb}{47}{11280}{1993}{EXACT
SOLUTION OF A QUANTUM MODEL OF RESONATING VALENCE BONDS ON THE
HEXAGONAL LATTICE}), using a string representation for the associated
dimer configurations.  Unfortunately there are inconsistencies
associated with Orland's string decomposition, and furthermore his
chosen boundary conditions on the hexagonal lattice do not accomodate
the uniform state. There is also a problem of overcounting associated
with the antisymmetrization procedure used, so the status of his
results of a gapless spectrum and no fluctuation-selection is
unclear. (C. L. Henley, S. Sachdev, N. Read, private communications).

\bibitem{henabs} This mapping was independently noted by C.L. Henley
in A. Gervois, M. Gingold and D. Iagolnitzer, eds.  {\sl Book of
Abstracts of STATPHYS 20 } (IUPAP Commission on Statistical Physics,
Paris, 1998).

\bibitem{jose} J. V. Jose, L. P. Kadanoff, S. Kirkpatrick and
D. R. Nelson, \tit{\prb}{16}{1217}{1977}{Renormalization, vortices,
and symmetry-breaking perturbatoins in the two-dimensional planar
model}.

\bibitem{mefvillain}
M. E. Fisher, \jour{\pr}{124}{1664}{1961}.

\bibitem{sachjala}
R. A. Jalabert and S. Sachdev,
\tit{\prb}{44}{686}{1991}{SPONTANEOUS ALIGNMENT OF FRUSTRATED 
BONDS IN AN ANISOTROPIC, 3-DIMENSIONAL ISING-MODEL}.

\bibitem{orlandsquare}
P. Orland, \tit{\prb}{49}{3423}{1994}{ROKHSAR-KIVELSON MODEL 
OF QUANTUM DIMERS AS A GAS OF FREE FERMIONIC STRINGS}

\bibitem{leung}
P. W. Leung, K. C. Chiu and K. J. Runge, 
\tit{\prb}{54}{12938}{1996}{Columnar dimer and plaquette 
resonating-valence-bond orders in the quantum dimer model}.

\bibitem{andersonpyro}
P. W. Anderson, \tit{\pr}{102}{1008}{1956}{Ordering and
Antiferromagnetism in Ferrites}

\bibitem{liebice}
E. H. Lieb, \jour{\pr}{162}{162}{1967}.

\bibitem{Liebmann86} For a review, see: R. Liebmann, {\sl Statistical
Mechanics of Periodic Frustrated Ising Systems}, (Springer, Berlin,
1986).

\bibitem{newbarkema} Different representations of the ice model are
explained in G. T. Barkema and M. E. J. Newman,
\tit{\pre}{57}{1155}{1998}{Monte Carlo simulation of ice models}

%\bibitem{sqcrclass} 
%E. H. Lieb and F. Y. Wu, in: Phase transitions and
%critical phenomena, Eds.: C. Domb and M. S. Green, Academic Press
%1972.

\bibitem{Kano53}
K. Kano and S. Naya, {\sl Prog. Theor. Phys. Japan}
{\bf 10}, 158 (1953).

\bibitem{Pauling38} 
L. Pauling, {\sl The Nature of the Chemical Bond}, (Cornell University
Press, Ithaca, 1938).

\bibitem{reimers}
J. N. Reimers, A. J. Berlinsky and A.-C. Shi,
\tit{\prb}{43}{865}{1991}{Mean-field approach to magnetic ordering in
highly frustrated pyrochlores}

\bibitem{Harris92} 
A.B. Harris, C. Kallin and A.J. Berlinsky, 
\jour{\prb}{45}{2899}{1992}.

\bibitem{suto} A. Suto, \tit{\zpb}{44}{121}{1981}{MODELS OF
SUPER-FRUSTRATION}; W. F. Wolff, P. Hoever and J. Zittartz,
\tit{\zpb}{42}{259}{1981}{LAYERED INHOMOGENEOUS ISING-MODELS 
WITH FRUSTRATION ON A SQUARE LATTICE}.

\bibitem{wolffzitt}
W. F. Wolff and J. Zittartz, \jour{\zpb}{44}{139}{1982}.

\end{references}
\end{document}